\documentclass[onecolumn,prd,superscriptaddress,nofootinbib,longbibliography,preprintnumbers]{revtex4-2}

\usepackage{hyperref}
\usepackage{graphicx}
\usepackage{svg}
\usepackage{float}
\usepackage{tabularx}
\usepackage{supertabular}
\usepackage{tablefootnote}
\usepackage{fancyhdr}
\usepackage{mathtools}
\usepackage{enumerate}
\usepackage{amsmath}
\usepackage{hhline}
\usepackage{amssymb,amsfonts}
\usepackage{amsthm}
\usepackage{physics}
\usepackage{subcaption}
\usepackage{orcidlink}

\usepackage{dcolumn}
\usepackage{bm}
\makeatletter
\let\@makefntextOrig\@makefntext
\def\@makefntext#1{\@makefntextOrig{\baselineskip=13pt #1}}
\makeatother
\begin{document}
\title{\boldmath Forward-mode automatic differentiation for the tensor renormalization group and its relation to the impurity method
}

\author{\orcidlinki{Yuto Sugimoto}{0009-0002-6735-3751}}
\email[E-mail: ]{sugimoto@nucl.phys.tohoku.ac.jp}
\affiliation{Department of Physics, Tohoku University, Sendai 980-8578, Japan}

\begin{abstract}
We propose a forward-mode automatic differentiation (AD) framework for tensor renormalization group methods. In this approach, evaluating the derivatives of the partition function up to the order of $k$ increases the matrix-multiplication cost by a factor of $(k+1)(k+2)/2$ compared to computing the free energy alone, and the memory footprint is only $k+1$ times that of the original calculation. 
In the limit where the derivatives of the singular value decomposition are neglected, we establish a theoretical correspondence between our forward-mode AD and conventional impurity methods.
Numerically, we find that the proposed AD algorithm can calculate internal energy and specific heat significantly higher accuracy than the impurity method at comparable computational cost.
We also provide a practical procedure to extract critical exponents from derivatives of the renormalized tensor in tensor renormalization group calculations in both two and three dimensions. 
In addition, we discuss how to efficiently differentiate an arbitrary tensor network.
\end{abstract}
\maketitle

\section{Introduction}
The tensor renormalization group (TRG)~\cite{levinTensorRenormalizationGroup2007} is a numerical method for studying classical and quantum many-body systems.
A wide range of applications has been developed, including tensor network formulations of path integrals for fermionic and bosonic theories as well as lattice gauge theories~\cite{guTensorEntanglementFilteringRenormalizationApproach2009,xieCoarsegrainingRenormalizationHigherorder2012a,shimizuCriticalBehaviorLattice2014a,unmuth-yockeyTensorRenormalizationGroup2014,yuTensorRenormalizationGroup2014,wangPhaseTransitionsFerromagnetic2014,shimizuGrassmannTensorRenormalization2014,takedaGrassmannTensorRenormalization2015a,takedaGrassmannTensorRenormalization2015,kawauchiTensorRenormalizationGroup2016,sakaiHigherOrderTensor2017,sakaiApplicationTensorNetwork2018,yoshimuraCalculationFermionicGreen2018,bazavovTensorRenormalizationGroup2019,iinoBoundaryTensorRenormalization2019,hongLogarithmicFinitesizeScaling2020,akiyamaPhaseTransitionFourdimensional2019,akiyamaTensorRenormalizationGroup2020,delcampComputingRenormalizationGroup2020,akiyamaTensorRenormalizationGroup2021,akiyamaRestorationChiralSymmetry2021,akiyamaPhaseTransitionFourdimensional2021,blochTensorRenormalizationGroup2021,fukumaTensorNetworkApproach2021,akiyamaMoreGrassmannTensor2021,hirasawaTensorRenormalizationGroup2021,akiyamaMetalInsulatorTransition2022,liTensornetworkRenormalizationApproach2022,nakayamaPhaseStructureCP12022,akiyamaTensorRenormalizationGroup2022,blochTensornetworkStudy3d2022,kuwaharaTensorRenormalizationGroup2022,luoTensorRenormalizationGroup2023,blochGrassmannHigherorderTensor2023,asaduzzamanImprovedCoarsegrainingMethods2023,yoshiyamaHigherorderTensorRenormalization2023,genzorCalculationCriticalExponents2023,akiyamaCriticalEndpoint3+1dimensional2023a,akiyama$SU2$PrincipalChiral2024,akiyamaTensorRenormalizationGroup2024,nakayamaInitialTensorConstruction2024b,samlodiaPhaseDiagramGeneralized2024,hommaTensorNetworkRenormalization2025,sugimotoPhaseStructure3+1dimensional2025,aizawaPhaseStructureAnalysis2025,kannoGrassmannTensorRenormalization2025}.
There have been several algorithmic developments aimed at improving computational efficiency~\cite{moritaTensorRenormalizationGroup2018,adachiAnisotropicTensorRenormalization2020,kadohRenormalizationGroupTriad2019} or achieving higher accuracy~\cite{adachiBondweightedTensorRenormalization2022,evenblyTensorNetworkRenormalization2015,yangLoopOptimizationTensor2017,hauruRenormalizationTensorNetworks2018,moritaGlobalOptimizationTensor2021,hommaNuclearNormRegularized2024,songVariationalBoundaryBased2025}.
In the TRG framework, the partition function is approximated via a truncation based on singular value decomposition (SVD), which retains the dominant contributions while discarding subleading degrees of freedom. 
Beyond the partition function, the computation of physical quantities is equally essential.
Three common approaches are available, numerical differentiation, the impurity-tensor method~\cite{moritaCalculationHigherorderMoments2019}, and automatic differentiation~\cite{liaoDifferentiableProgrammingTensor2019}. However, numerical differentiation based on finite differences often suffers from numerical instability, since its accuracy strongly depends on the choice of the step size.
The impurity method evaluates observables by inserting an impurity-tensor into the tensor network. Depending on the target quantity, this approach yields smoother data than numerical differentiation. Extensions of the method allow for systematic summations that directly access higher-order moments, enabling the evaluation of quantities such as the Binder cumulant or magnetization for the higher-order TRG (HOTRG) and bond-weighted TRG (BWTRG)~\cite{moritaCalculationHigherorderMoments2019,moritaMultiimpurityMethodBondweighted2024}. However, in the impurity method, one typically employs the same projectors as those optimized for the bulk tensor which can be an additional source of systematic error.
Automatic differentiation (AD) ~\cite{bartholomew-biggsAutomaticDifferentiationAlgorithms2000} offers a powerful alternative for differentiation, enabling machine precision evaluation of derivatives. AD treats the computational process as a graph, iteratively applying the chain rule to propagate derivatives. This methodology has proven highly successful in the field of deep learning~\cite{baydinAutomaticDifferentiationMachine2018}.
Notably, Ref.~\cite{liaoDifferentiableProgrammingTensor2019} demonstrated a straightforward extension of AD to tensor network algorithms including TRG by interpreting the tensor network computation as a differentiable computational graph.
Several studies have explored the incorporation of AD techniques within the context of the TRG~\cite{chenAutomaticDifferentiationSecond2020,gengDifferentiableProgrammingIsometric2022,adachiBondweightedTensorRenormalization2022,genzorCalculationCriticalExponents2023,tsengBondWeightedTensorRenormalization2025}. 
Ref.~\cite{liaoDifferentiableProgrammingTensor2019} uses backpropagation (reverse-mode AD), which evaluates gradients of a given target function via a backward pass, to compute the internal energy and the specific heat within the TRG framework.
While reverse-mode AD is well suited to problems with many input parameters such as neural networks, its memory cost typically scales with the depth of the computational graph, because it requires storing intermediate quantities for the backward pass that propagates sensitivities from the output back to the inputs. 
In the TRG setting, the depth of the computational graph grows logarithmically with the system size, hence, the reverse-mode memory footprint increases accordingly.
This becomes particularly restrictive when one requires large volumes or considers higher-dimensional tensor networks, where even the storage of individual tensors is already demanding.
Moreover, the TRG scheme is typically designed to evaluate the partition function over a range of system sizes within a single forward coarse-graining procedure.
In such cases, the additional memory and computational overhead associated with backpropagation can be prohibitive, especially when analyzing the volume dependence of observables, for example in finite-size scaling studies.
Furthermore, existing implementations in ~\cite{tensorgrad_github,tensornetworkadjl_github} rely on high-level AD packages~\cite{paszkePyTorchImperativeStyle2019,Zygote.jl-2018}, which may limit their applicability to specific programming languages or environments.

Motivated by these limitations, we develop an alternative strategy to incorporate automatic differentiation into the TRG scheme by explicitly deriving the forward-mode chain rule. 
Our approach can be viewed as a generalization of the impurity method. It propagates derivative information along the renormalization flow in a controlled manner, without requiring backpropagation through a deep computational graph. Consequently, both the memory footprint and the computational cost remain only a small constant factor larger than those of the original TRG computation. The proposed method also achieves substantially higher accuracy while preserving the same asymptotic computational scaling as conventional impurity-based calculations in the HOTRG. Furthermore, the method is readily applicable to a broad class of TRG algorithms and can be extended straightforwardly to higher-dimensional tensor networks. Finally, we present an efficient procedure to extract critical exponents by applying the proposed AD scheme to finite-size scaling analyses.

In Sec.~\ref{sec:method}, we review automatic differentiation and present our proposed algorithm, together with its theoretical connection to the impurity method of the HOTRG. We also extend our scheme to the BWTRG~\cite{adachiBondweightedTensorRenormalization2022}. In Sec.~\ref{sec:results}, we provide numerical results that demonstrate the efficiency of the proposed approach. We show that it yields significantly more accurate estimates while preserving the same order of computational cost as the conventional impurity method. We further perform a finite-size scaling analysis enabled by our AD-based evaluation, and present additional results for the three-dimensional isotropic lattice case. Sec.~\ref{sec:summary} is devoted to discussion and summary.
\section{Methods}\label{sec:method}
\subsection{About tensor renormalization group}
We first briefly review the tensor renormalization group method~\cite{levinTensorRenormalizationGroup2007,xieCoarsegrainingRenormalizationHigherorder2012a}.
For a translationally invariant system, the partition function $Z$ in two dimensions can be written as a uniform tensor network,
\begin{equation}\label{eq:tensor_Z}
  Z=\mathrm{tTr}\left[\,\vcenter{\hbox{\includegraphics[scale=0.55]{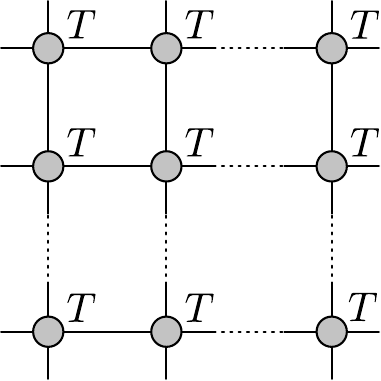}}}\,\right],
\end{equation}
where $\mathrm{tTr}$ denotes the contraction of all tensor indices with periodic boundary conditions. The local tensor $T$ derived from the action is placed on each lattice site and connected according to the square (hypercubic) lattice geometry. However, a direct contraction of the tensor network representation of $Z$ is impractical because the computational cost grows exponentially with the system volume $V$.

TRG methods address this difficulty by employing coarse-graining of the network, typically using a truncated singular value decomposition to control the bond dimension. In TRG schemes such as the Levin--Nave-type TRG~\cite{levinTensorRenormalizationGroup2007,evenblyTensorNetworkRenormalization2015,adachiBondweightedTensorRenormalization2022}, or the HOTRG type~\cite{xieCoarsegrainingRenormalizationHigherorder2012a,kadohRenormalizationGroupTriad2019,adachiAnisotropicTensorRenormalization2020}, the local tensor $T$ is approximated or decomposed and contracted with its neighbors recursively, so that the number of tensors in the network decreases at each iteration. Importantly, for a translationally invariant system, the network remains uniform under coarse-graining; namely the same tensor or unit cell is retained on every coarse-grained lattice site.

Typically, the coarse-graining map $R$ for the HOTRG on a two-dimensional lattice along the $x$ direction is based on the contraction:
\begin{align}\label{eq:HOTRG}
  \vcenter{\hbox{\includegraphics[scale=0.7]{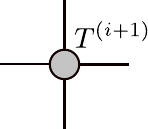}}}
  &\coloneqq
  R\bigl(T^{(i)},T^{(i)},E^{(i)},F^{(i)}\bigr)\nonumber\\
  &=\vcenter{\hbox{\includegraphics[scale=0.65]{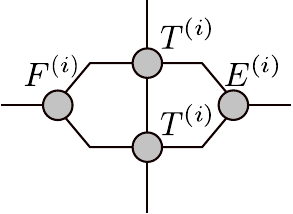}}}.
\end{align}
Here, $T^{(i)}$ and $T^{(i+1)}$ denote the tensors before and after coarse-graining at steps $i$ and $i+1$, respectively, and $E^{(i)}$ and $F^{(i)}$ are projectors inserted into every horizontal link of lattice, which effectively performs a truncated SVD of $T^{(i)}$ or a larger subnetwork.
In the above contraction, two adjacent tensors $T^{(i)}$ are approximated into a single tensor $T^{(i+1)}$, which we call coarse-grained tensor.
By iterating the above procedure $O(\log V)$ times, the original network is reduced to a single tensor, and the approximated partition function under the periodic boundary condition is obtained by taking its tensor trace.

\subsection{Automatic differentiation and tensor renormalization group}
Consider a scalar function
\begin{align}
f(\theta_1,\ldots,\theta_n),\qquad \boldsymbol{\theta}\in\mathbb{R}^n,
\end{align}
where $\boldsymbol{\theta}=(\theta_1,\ldots,\theta_n)^\mathsf{T}$.
We assume that $f$ is given as a composition of sequentially defined maps
\begin{align}
f(\boldsymbol{\theta}) = \left(w_J \circ w_{J-1} \circ \cdots \circ w_0\right)(\boldsymbol{\theta}).
\end{align}
Here, $f$ denotes a certain algorithm, while each $w_k$ represents a computational routine that maps an intermediate variable to the next one. By introducing intermediate variables $\{x_k\}_{k=0}^{J}$ as
\begin{align}
x_0 &= \boldsymbol{\theta},\\
x_{j+1} &= w_j(x_j)\qquad (j=0,1,\dots,J),
\end{align}
where $x_{J+1}=f$.
Our goal is to evaluate the derivatives of $f$ with respect to the parameters $\{\theta_i\}_{i=1}^n$. 
By applying the chain rule, we obtain
\begin{align}\label{eq:chain_rule}
\pdv{f}{\theta_i}&=\pdv{f}{x_{J+1}}\pdv{x_{J+1}}{x_{J}}\cdots\pdv{x_0}{\theta_i}.
\end{align}
AD computes these derivatives exactly up to machine precision by decomposing $f$ into a computational graph composed of the elementary maps $\{w_j\}$ and automatically evaluating the derivatives through the chain rule. 
The forward-mode AD evaluates Eq.~\eqref{eq:chain_rule} from right to left along the computational graph:
\begin{align}
\pdv{x_j}{\theta_i}
=
\pdv{x_j}{x_{j-1}}
\pdv{x_{j-1}}{\theta_i},
\end{align}
for $j=1,\dots,J+1$. This strategy is useful when the number of input parameters is small while the output $f$ is high dimensional, such as a vector or a tensor.

Another strategy evaluates Eq.~\eqref{eq:chain_rule} from left to right. This is called reverse-mode AD. Defining the adjoint variables
\begin{align}
\bar{x}_j \equiv \pdv{f}{x_j},
\end{align}
reverse-mode AD propagates them backward as
\begin{align}
\bar{x}_{j-1} = \bar{x}_j \pdv{x_j}{x_{j-1}},
\end{align}
for $j=J+1,\dots,1$, yielding $\bar{x}_0=\pdv{f}{\theta_i}$.
Since we do not have to distinguish the dependence on each $\theta_i$ until the final step, we can compute the derivatives with respect to $\theta_i$ for $i=1,\dots,n$ in a single backward pass. This is particularly effective when the number of input parameters is much larger than the dimension of the output, as in neural networks.

Let us discuss the TRG case. Throughout this paper, we assume that the initial tensor $T^{(0)}$ depends on the parameter $\beta$. The function $f$ corresponds to the free energy or the partition function of the system, and the intermediate variables $x_j$ correspond to the renormalized tensors $T^{(j)}$. In Ref.~\cite{liaoDifferentiableProgrammingTensor2019}, a reverse-mode implementation was introduced to compute the internal energy and the specific heat for the two-dimensional Ising model. However, reverse-mode AD faces practical difficulties in TRG calculations. Each TRG step relies on a truncated SVD of $T^{(j)}$, so the map $w_j$ depends strongly on $T^{(j)}$ through the associated projectors. Consequently, to carry out the backward pass one has to store all projectors, or the tensors $T^{(j)}$ generated at each renormalization step. This leads to significant memory and disk usage.

Another limitation is that reverse-mode AD computes the gradient of a specific target function. 
Denoting one TRG iteration by
\begin{equation}\label{eq:R}
T^{(i+1)} = R\left(T^{(i)},E_1^{(i)},E_2^{(i)},\ldots\right).
\end{equation}

Here, $\{E_k^{(i)}\}$ denotes the projectors used in the $i$th step.
Assuming that the renormalization map $R$ increases the system size by a factor of two, the partition function for different system sizes can be written as

\begin{align}
Z(V=2^1,\beta) &= \mathrm{tTr}\left[R\left(T^{(0)}(\beta)\right)\right], \nonumber\\
Z(V=2^2,\beta) &= \mathrm{tTr}\left[(R \circ R)\left(T^{(0)}(\beta)\right)\right], \nonumber\\
&\vdots \nonumber\\
Z(V=2^n,\beta) &= \mathrm{tTr}\left[\underbrace{R \circ \cdots \circ R}_{n}\!\left(T^{(0)}(\beta)\right)\right].\label{eq:R_V}
\end{align}
To obtain derivatives of $Z$ for arbitrary system sizes with reverse-mode AD, one would, in principle, need to apply reverse-mode AD separately to $R$, $R \circ R$, \dots, and $\underbrace{R \circ \cdots \circ R}_{n}$. This is inefficient, because a single TRG run naturally produces the sequence of tensors $\{T^{(i)}\}$, and, thus, all system sizes in Eq.~\eqref{eq:R_V}, within one calculation.
Furthermore, the renormalized tensor $T^{(i)}$ itself contains physical information beyond the partition function, such as scaling dimensions~\cite{guTensorEntanglementFilteringRenormalizationApproach2009} and energy spectrum~\cite{az-zahraSpectroscopyTensorRenormalization2024}, so TRG should be viewed as a multioutput algorithm whose outputs are the entire sequence $\{T^{(i)}\}$.
From this perspective, the forward-mode AD is particularly suitable, since it propagates derivative information together with the forward renormalization flow and can provide derivatives associated with all outputs $\{T^{(i)}\}$ in a single run.
\subsection{Differentiation of the HOTRG}
We introduce the forward-mode AD of the HOTRG here.
First, we consider first derivatives of the partition function $Z$.
The partition function $Z^{(n)}$ at $n$th iteration is defined as,
\begin{equation}
  Z^{(n)}=\mathrm{tTr}[T^{(n)}].
\end{equation}
Assume the initial tensor or action has parameter dependence with respect to $\beta$.
The first derivative of $\ln Z$ with respect to $\beta$ at the $n$th iteration is simply given by
\begin{equation}
  \pdv{\ln Z^{(n)}}{\beta}
  =\frac{\mathrm{tTr}\left[\pdv{T^{(n)}}{\beta}\right]}{\mathrm{tTr}[T^{(n)}]}
  \label{eq:first_derivative}
\end{equation}
Therefore, the entire procedure requires computing only two objects at each renormalization step: the coarse-grained tensor $T^{(n)}$ and its derivative $\dot{T}^{(n)}\coloneqq \pdv{T^{(n)}}{\beta}$.

To derive the chain rule, let us consider the RG map of 2D HOTRG in Eq.~\eqref{eq:HOTRG}.
To calculate $\dot{T}^{(n)}$ at each step, we begin with the initial tensor $T^{(0)}$ and its derivative $\dot{T}^{(0)}$.
The chain rule for the HOTRG update is expressed as the derivative of RG map $R$. By differentiating Eq.\eqref{eq:HOTRG} from $i=1,\dots,n$, sequentially, we get
\begin{align}
  \dot{T}^{(n+1)}
  &=
  R\bigl(\dot{T}^{(n)}, T^{(n)}, E^{(n)}, F^{(n)}\bigr)
  + R\bigl(T^{(n)}, \dot{T}^{(n)}, E^{(n)}, F^{(n)}\bigr) \nonumber\\
  &\quad
  + R\bigl(T^{(n)}, T^{(n)}, \dot{E}^{(n)}, F^{(n)}\bigr)
  + R\bigl(T^{(n)}, T^{(n)}, E^{(n)}, \dot{F}^{(n)}\bigr).
  \label{eq:first_chain_rule}
\end{align}
The derivatives of $E$ and $F$ can be obtained through a differentiation of the SVD formula, which we leave for Appendix~\eqref{ap:SVD}. In the practical calculation, we adopt a Lorentzian broadening $1/x\rightarrow x/(x^2+\eta)$ for derivatives of isometries following Ref.~\cite{liaoDifferentiableProgrammingTensor2019} to avoid divergence when there are degenerate singular values. 
In this work, we will use improved projectors refereed as squeezers~\cite{wangClusterUpdateTensor2011,iinoBoundaryConformalSpectrum2020,adachiAnisotropicTensorRenormalization2020} in the HOTRG calculations. In the Ising model which we use in this work, $\eta\rightarrow \infty$ limit corresponds to the vanishing limit of the derivatives of squeezers\footnote{This is because initial tensor of Ising model has reflection symmetry, which yields squeezers to be isometry itself. In general squeezers, $\eta\rightarrow\infty$ does not mean $\dot{E}=\dot{F}=0$ without taking derivatives of singular values to be zero.}.
The above procedure completes the formulation of the first-order response within the forward-mode scheme.
The extension to second-order derivatives is completely analogous.
By differentiating Eq.~\eqref{eq:first_chain_rule} once more with respect to $\beta$,
we obtain the evolution equation for the second derivative $\ddot{T}^{(n+1)}$ as follows with $T^{(n)}$, $\dot{T}^{(n)}$, and $\ddot{T}^{(n)}$:
\begin{align}
  \ddot{T}^{(n+1)}
  &=
  R\bigl(\ddot{T}^{(n)}, T^{(n)}, E^{(n)}, F^{(n)}\bigr)
  + R\bigl(T^{(n)}, \ddot{T}^{(n)}, E^{(n)}, F^{(n)}\bigr) \nonumber\\
  &\quad
  + R\bigl(T^{(n)}, T^{(n)}, \ddot{E}^{(n)}, F^{(n)}\bigr)
  + R\bigl(T^{(n)}, T^{(n)}, E^{(n)}, \ddot{F}^{(n)}\bigr) \nonumber\\
  &\quad
  + 2\Bigl[R\bigl(\dot{T}^{(n)}, \dot{T}^{(n)}, E^{(n)}, F^{(n)}\bigr)
    + R\bigl(T^{(n)}, T^{(n)}, \dot{E}^{(n)}, \dot{F}^{(n)}\bigr).
    \nonumber\\
    &\quad
    + R\bigl(\dot{T}^{(n)}, T^{(n)}, \dot{E}^{(n)}, F^{(n)}\bigr)
    + R\bigl(\dot{T}^{(n)}, T^{(n)}, E^{(n)}, \dot{F}^{(n)}\bigr)
    \nonumber\\
    &\quad
    + R\bigl(T^{(n)}, \dot{T}^{(n)}, \dot{E}^{(n)}, F^{(n)}\bigr)
  + R\bigl(T^{(n)}, \dot{T}^{(n)}, E^{(n)}, \dot{F}^{(n)}\bigr)\Bigr].
  \label{eq:second_chain_rule}
\end{align}
Taking the derivative of Eq.~\eqref{eq:first_derivative} with respect to $\beta$, we obtain
\begin{equation}
  \pdv[2]{\ln Z^{(n)}}{\beta}=\frac{\mathrm{tTr}\left[\ddot{T}^{(n)}\right]\ \mathrm{tTr}\left[T^{(n)}\right]
  -\left(\mathrm{tTr}\left[\dot{T}^{(n)}\right]\right)^{2}}
  {\left(\mathrm{tTr}\left[T^{(n)}\right]\right)^{2}}.
  \label{eq:second_derivative}
\end{equation}
Higher-order derivatives follow straightforwardly by iterating the same procedure.
The update rule for $\dot T$ in Eq.~\eqref{eq:first_chain_rule} naively consists of four HOTRG-like contractions. However, we do not need to evaluate these four contractions separately. Instead, we use the fact that each contraction can be decomposed into a sequence of matrix-multiplications along a contraction-tree~\cite{evenblyImprovingEfficiencyVariational2014}.
Rather than evaluating the four contractions independently, we differentiate the intermediate tensors that appear along this tree. With this procedure, the total matrix multiplication cost of computing derivatives up to first order is exactly 3 times that of the original HOTRG.

The same idea extends to higher-orders. Although the algorithm involves a series of HOTRG-like contractions, contraction-tree-based differentiation reduces the total numerical cost of evaluating all derivatives up to the order of $k$ to only $\frac{(k+1)(k+2)}{2}$ times that of computing the free energy alone.
For example, by explicitly rewriting Eq.~\eqref{eq:first_chain_rule} as
\begin{align}\label{eq:chain rule_HOTRG}
  \vcenter{\hbox{\includegraphics[scale=0.7]{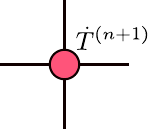}}}&=\vcenter{\hbox{\includegraphics[width=0.75\linewidth]{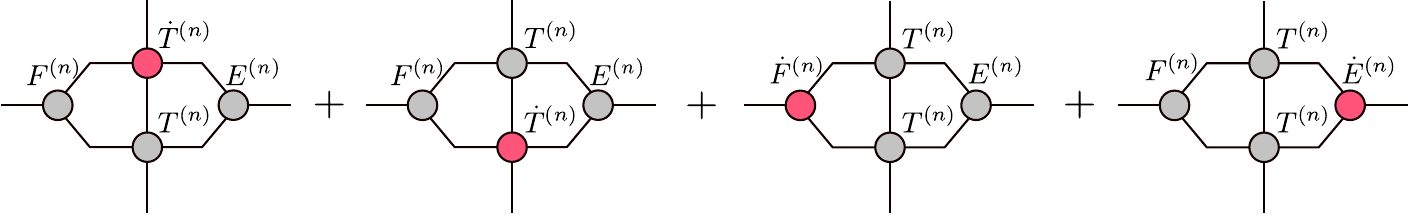}}}\nonumber\\
  &\quad=\vcenter{\hbox{\includegraphics[width=0.75\linewidth]{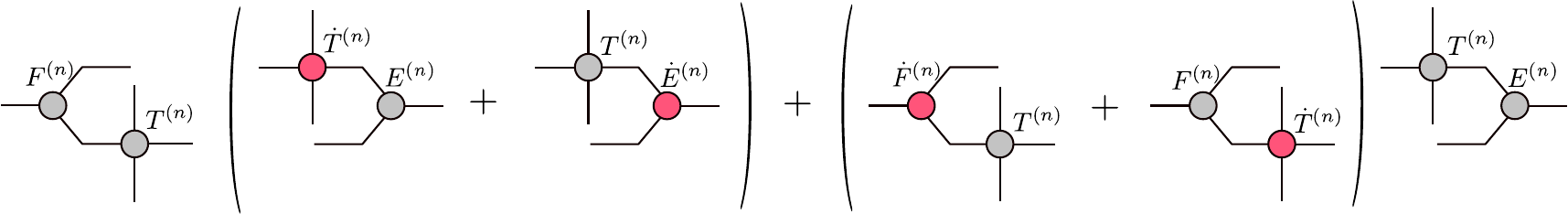}}},
\end{align}
where the red tensor represents first derivative tensor.
In Eq.~\eqref{eq:chain rule_HOTRG}, the dominant matrix-multiplication part of the contraction is evaluated only twice.

For the second derivative, Eq.~\eqref{eq:second_chain_rule} can be written as
\begin{align}\label{eq:chain rule2_HOTRG}
  \vcenter{\hbox{\includegraphics[scale=0.7]{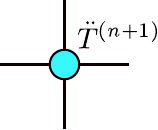}}}&=\vcenter{\hbox{\includegraphics[width=0.65\linewidth]{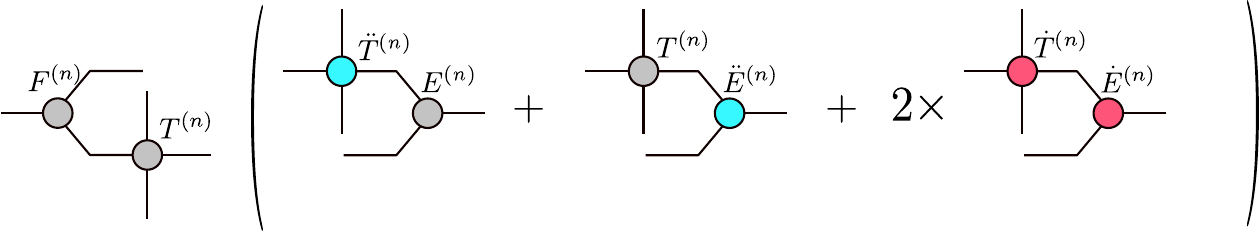}}}\nonumber\\
  &\quad+\quad\vcenter{\hbox{\includegraphics[width=0.65\linewidth]{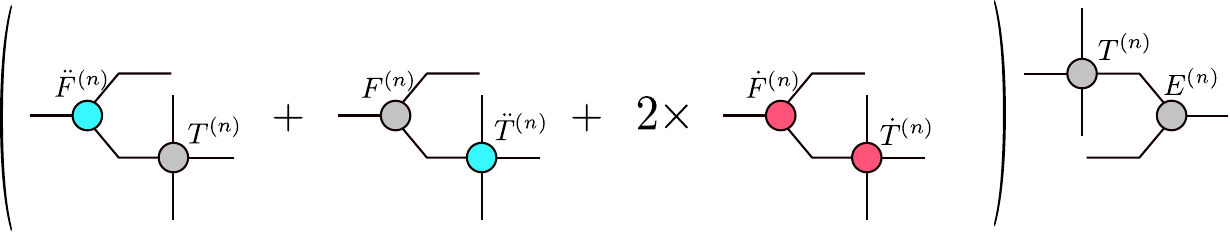}}}\nonumber\\
  &\quad+\quad\vcenter{\hbox{\includegraphics[width=0.65\linewidth]{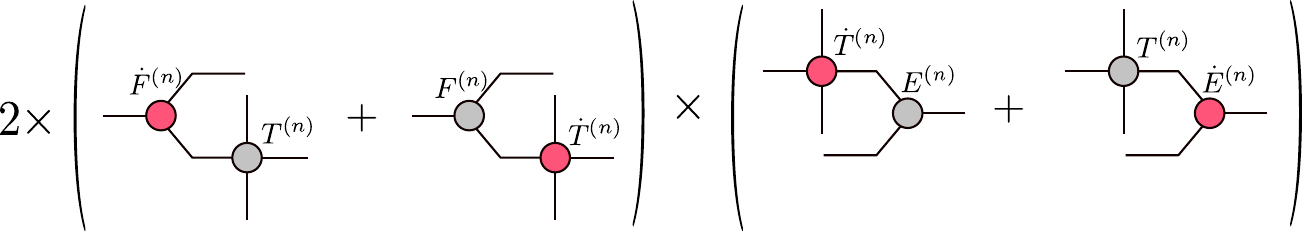}}}\quad.
\end{align}
The blue tensors indicate its second derivative.
In this contraction, the bottleneck takes just 3 times as the HOTRG.
In practice, it is not necessary to execute all contractions shown in the figures independently. Instead, by performing the bulk, first-order, and second-order contractions sequentially for each intermediate tensor, one can reuse the intermediate tensors that are common to Eqs.~\eqref{eq:HOTRG}, \eqref{eq:chain rule_HOTRG}, and \eqref{eq:chain rule2_HOTRG}, rather than recomputing them multiple times. 
As a result, the total cost of evaluating derivatives up to first and second order is exactly 3 and 6 times that of the original HOTRG, respectively. This scaling is unchanged for general tensor network contractions. See Appendix~\ref{ap:general_case} for details. 
We also note that the memory footprint is only $k+1$ times that of the original HOTRG, since we only propagate $T^{(i)}$ and its derivatives at every coarse-graining step; importantly, it does not depend on the graph depth $\log V$.
Note that, to obtain the derivatives of the squeezers (isometries), we differentiate the SVD and apply the resulting rules to all contractions appearing in the original algorithm.
This approach is applicable to a wide range of programming languages without relying on external AD packages, as it requires only implementing a simple update rule.
\subsection{Differentiation of the BWTRG}
Here we employ AD rules for the BWTRG~\cite{adachiBondweightedTensorRenormalization2022}.
Suppose we have fundamental tensor $T^{(n)}$ and tensor $A^{(n)},B^{(n)},C^{(n)},D^{(n)}$ constructed from SVD of $T^{(n)}$ and bond weight $E^{(n)},F^{(n)}$.
In the coarse-graining step of BWTRG, we consider the following contraction:
\begin{equation}
  \vcenter{\hbox{\includegraphics[scale=1.0]{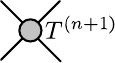}}}\coloneqq\vcenter{\hbox{\includegraphics[scale=0.9]{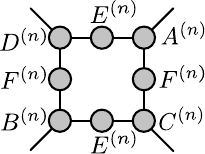}}}
\end{equation}
By considering the SVD of $T^{(n)}$ and its derivative, we can easily obtain $\dot{A}^{(n)},\dot{B}^{(n)},\dot{C}^{(n)},\dot{D}^{(n)}$ or higher-order derivatives.
Before the contraction, we introduce the following intermediate tensor,
\begin{align}
  \vcenter{\hbox{\includegraphics[scale=1.0]{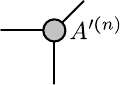}}}&=\vcenter{\hbox{\includegraphics[scale=1.0]{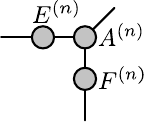}}}\\
  \vcenter{\hbox{\includegraphics[scale=1.0]{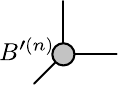}}}&=\vcenter{\hbox{\includegraphics[scale=1.0]{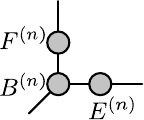}}}\\
\end{align}
and their derivative $\dot{A}',\dot{B}'$ by chain rule.
The coarse-graining rule for the BWTRG is now written as
\begin{align}\label{eq:BWTRG}
  \vcenter{\hbox{\includegraphics[scale=1.0]{fig_old/T_n1_BTRG.pdf}}}&\coloneqq R_{\text{BWTRG}}(A',B',C^{(n)},D^{(n)})\nonumber\\
  &=\vcenter{\hbox{\includegraphics[scale=0.9]{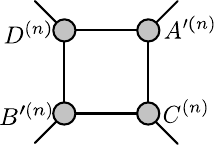}}}.
\end{align}The first derivatives can be obtained by differentiating Eq.~\eqref{eq:BWTRG},
\begin{align}\label{eq:BWTRG_d1}
  \vcenter{\hbox{\includegraphics[scale=1.1]{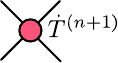}}}&= R_{\text{BWTRG}}(\dot{A'},B',C^{(n)},D^{(n)})+R_{\text{BWTRG}}(A',\dot{B}',C^{(n)},D^{(n)})\nonumber\\
  &\qquad+R_{\text{BWTRG}}(A',B',\dot{C}^{(n)},D^{(n)})+R_{\text{BWTRG}}(A',B',C^{(n)},\dot{D}^{(n)})\\
  &=\vcenter{\hbox{\includegraphics[width=0.7\linewidth]{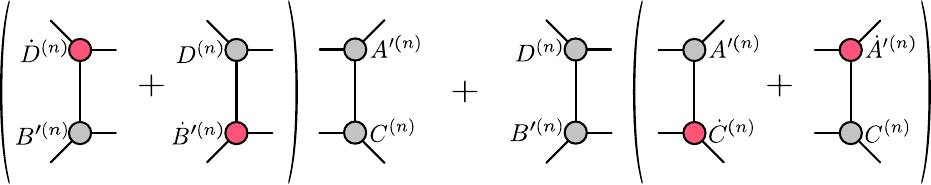}}}
\end{align}

\begin{align}\label{eq:BWTRG_d2}
  \vcenter{\hbox{\includegraphics[scale=1.1]{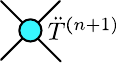}}}&= R_{\mathrm{BWTRG}}\left(\ddot{A}',B',C^{(n)},D^{(n)}\right)
  +R_{\mathrm{BWTRG}}\left(A',\ddot{B}',C^{(n)},D^{(n)}\right)\nonumber\\
  &\quad+R_{\mathrm{BWTRG}}\left(A',B',\ddot{C}^{(n)},D^{(n)}\right)
  +R_{\mathrm{BWTRG}}\left(A',B',C^{(n)},\ddot{D}^{(n)}\right)
  \nonumber\\
  &\quad
  +2\Bigl[
    R_{\mathrm{BWTRG}}\left(\dot{A}',\dot{B}',C^{(n)},D^{(n)}\right)
    +R_{\mathrm{BWTRG}}\left(\dot{A}',B',\dot{C}^{(n)},D^{(n)}\right)
    +R_{\mathrm{BWTRG}}\left(\dot{A}',B',C^{(n)},\dot{D}^{(n)}\right)
    \nonumber\\
    &\qquad\quad
    +R_{\mathrm{BWTRG}}\left(A',\dot{B}',\dot{C}^{(n)},D^{(n)}\right)
    +R_{\mathrm{BWTRG}}\left(A',\dot{B}',C^{(n)},\dot{D}^{(n)}\right)
    +R_{\mathrm{BWTRG}}\left(A',B',\dot{C}^{(n)},\dot{D}^{(n)}\right)
  \Bigr]\\
  &=\vcenter{\hbox{\includegraphics[scale=0.8]{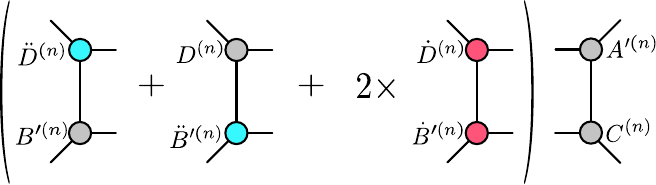}}}\nonumber\\
  &\quad+\vcenter{\hbox{\includegraphics[scale=0.8]{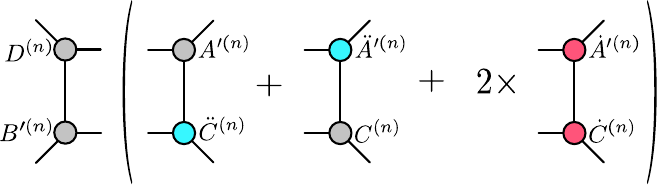}}}\nonumber\\
  &\quad+\vcenter{\hbox{\includegraphics[scale=0.8]{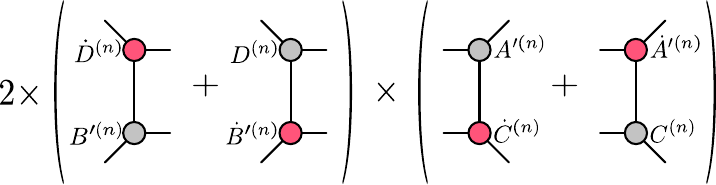}}}\,.
\end{align}
Again, the cost of computing derivatives up to second order is exactly 6 times the cost of the original contraction.
More generally, when derivatives up to the order of $k$ are propagated, the leading contraction cost is multiplied by $(k+1)(k+2)/2$.
In practical calculations, differentiating the SVD introduces an additional $O(D^6)$ overhead, while the remaining tensor contraction operations required to evaluate the derivatives are also multiplied by $(k+1)(k+2)/2$.
\subsection{A connection to impurity method}
Here we explain how the forward-AD scheme connects to impurity methods.
Let us consider the case evaluating the first derivatives of $\ln{Z}$. This value can be expressed as the summation of all possible configurations including a single $\dot{T}$~\cite{nakayamaInitialTensorConstruction2024},
\begin{align}\label{eq:imp_U}
  \pdv{Z}{\beta}=\mathrm{tTr}\left[\,\vcenter{\hbox{\includegraphics[scale=0.45]{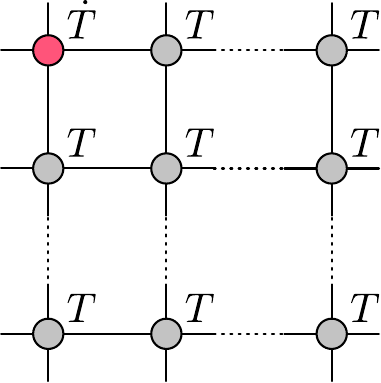}}}\,+\,\vcenter{\hbox{\includegraphics[scale=0.45]{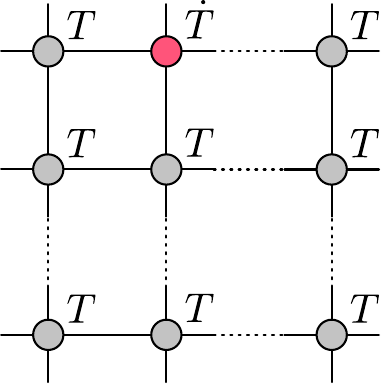}}}\,+\dots +\,\vcenter{\hbox{\includegraphics[scale=0.45]{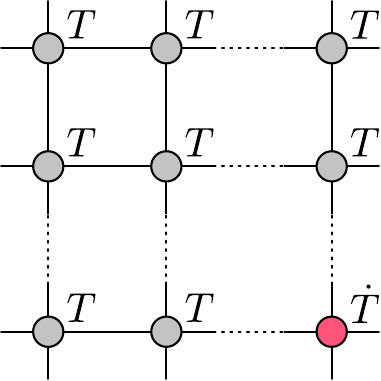}}}\,\right]
\end{align}
where $\dot{T}$ is called a single impurity-tensor. If there is translational invariance, Eq.~\eqref{eq:imp_U} can be treated as a single-impurity-tensor network localized at a specific lattice site. More generally, we can consider higher-order derivatives or operator insertions. In both cases, the impurity-tensor network can be systematically generated by introducing external sources and the corresponding generating function. The second derivative of $Z$ is also expressed as a summation of all possible networks containing any two $\dot{T}$ and the sum of networks containing one second derivative, $\ddot{T}$, which is called a two-impurity-tensor.

In the impurity method, tensor networks with impurities are coarse-grained by a given TRG scheme. There are two typical strategies: (i) use the same isometries as those obtained from the bulk tensor $T$, and (ii) introduce an independent truncation for the impurity-tensor itself. The former approach is often employed in HOTRG calculations of higher-order moments of physical observables and can be straightforwardly extended to higher-dimensional lattices. The latter approach is mainly practical in two-dimensional systems, because under TRG iterations impurities tend to spread over higher-dimensional lattices and for multiple impurities it becomes difficult to keep track of and sum over all possible impurity configurations.
Therefore, we will examine the former case here.

Let us now discuss the relationship between our forward-AD scheme and the conventional impurity method.
\subsubsection{Impurity method of the HOTRG}
In the impurity method in the HOTRG, $k$th impurity-tensor is updated as systematic summation of networks~\cite{moritaCalculationHigherorderMoments2019,yoshiyamaHigherorderTensorRenormalization2023},
\begin{align}
  S_1^{(n+1)}
  &=\frac{1}{2}\left[R\bigl(S_1^{(n)}, T^{(n)}, E^{(n)}, F^{(n)}\bigr)
  + R\bigl(T^{(n)}, S_1^{(n)}, E^{(n)}, F^{(n)}\bigr)\right]\label{eq:first_imp_rule}\\
  S_2^{(n+1)}
  &=\frac{1}{4}\Bigl[R\bigl(S_2^{(n)}, T^{(n)}, E^{(n)}, F^{(n)}\bigr)
  + R\bigl(T^{(n)}, S_2^{(n)}, E^{(n)}, F^{(n)}\bigr) + 2R\bigl(S_1^{(n)}, S_1^{(n)}, E^{(n)}, F^{(n)}\bigr)\Bigr],\label{eq:second_imp_rule}\\
  \vdots\\
  S_k^{(n+1)}
  &=\frac{1}{2^k}\sum_{m=0}^{k}\binom{k}{m} R\left(S_{k-m}^{(n)},S_{m}^{(n)},E^{(n)}, F^{(n)}\right).
  \label{eq:general_imp_rule}
\end{align}
where $S_m^{(n)}$ denote the $k$-impurity-tensors, with the initial conditions
\begin{align}
  S_m^{(0)}=\pdv[m]{T}{\beta}
\end{align}
and we have defined $S_0^{(i)}=T^{(i)}$ for all $i$. Thus, we can calculate up to $k$th derivatives at the cost of $\frac{(k+1)(k+2)}{2}$ times bottleneck contraction in the HOTRG.
The key observation is that the usual impurity method can be regarded as the limit of our forward-mode AD formulation in which the derivatives of the SVD are set to zero up to a normalization constant. Let us look at Eq.~\eqref{eq:first_chain_rule}. By setting $\dot{E} = \dot{F} = 0$, or $\eta=\infty$, we recover the HOTRG contraction with a single derivative tensor, where the projectors are those obtained from the bulk tensor $T$. This is essentially the standard impurity-tensor update scheme in the HOTRG method, once we identify $\dot{T}$ with an impurity-tensor. Likewise, for the second derivative, by interpreting $\ddot{T}$ as a two-impurity-tensor, we obtain an analogous correspondence. As discussed above, since an impurity-tensor can be expressed in terms of derivatives of the initial tensor, the forward-mode AD formulation can be regarded as an extension of the impurity method. This relationship naturally extends to higher-order derivatives. Therefore, even for higher-order derivatives, one can directly derive the update rules of the impurity-tensor method corresponding to arbitrary higher-order moments by applying the chain rule step by step, without explicitly accounting for the systematic summation.

\subsubsection{Impurity method of the BWTRG}\label{subsec:BWTRG}
To clarify the correspondence with the multi-impurity method for the BWTRG~\cite{moritaMultiimpurityMethodBondweighted2024}, we rewrite the BWTRG and its differentiation rules in a projector formalism~\cite{evenblyAlgorithmsTensorNetwork2017,nakamuraTensorRenormalizationGroup2019}.
This can be achieved by a simple modification: Replace the truncated SVD and the bond-weight splitting by an insertion of projectors and singular values,
\begin{align}
T^{(n)}
&\approx U\,\Sigma^{\frac{1-k}{2}}\,\Sigma^{k}\,\Sigma^{\frac{1-k}{2}}\,V^{\dagger}
\\
&=
U\,\Sigma^{\frac{1-k}{2}}
\left(
\Sigma^{-\frac{1-k}{2}}\,U^{\dagger}\,T^{(n)}\,V\,\Sigma^{-\frac{1-k}{2}}
\right)
\Sigma^{\frac{1-k}{2}}\,V^{\dagger},
\end{align}
where $k$ denotes the hyperparameter of the BWTRG, and the projectors $U,\,V^{\dagger}$ and $\Sigma$ are constructed from the truncated singular vectors and singular values of $T^{(n)}$, and the term inside the parentheses plays the role of the bond weight $E^{(n)}$ or $F^{(n)}$.
As in the HOTRG case, in the limit where the SVD differentiation is ignored and only the derivative of $T^{(n)}$ is retained, the chain rule in Eqs.~\eqref{eq:BWTRG} and \eqref{eq:BWTRG_d2} reproduces the update rule for the squared-lattice version of the multi-impurity method introduced in Ref.~\cite{moritaMultiimpurityMethodBondweighted2024} up to a normalization constant.
In this impurity limit, the derivatives of $E^{(n)}$ and $F^{(n)}$ play the role of spatial averages of higher-order derivatives.
Note that since the derivative of $T^{(n)}$ does not appear in the update rule in the BWTRG, the initial tensor in Eq.~\eqref{eq:imp_U} is inadequate in this limit.
Therefore, we must incorporate the temperature dependence into the initial bond weight via diagonalization of the Boltzmann weight and take $T^{(0)}$ to be constant with respect to $\beta$.
We also note that, once derivatives of the SVD are neglected, the derivatives of $E^{(n)}$ and $F^{(n)}$ are no longer constrained to be diagonal matrices.
The difference between AD and the impurity method was already pointed out in Ref.~\cite{moritaMultiimpurityMethodBondweighted2024}, where the impurity method effectively ignores the $\beta$ dependence of the isometric tensors.
In our formulation, this distinction becomes clearer.
Finally, we emphasize that even if a systematic impurity method cannot be defined, our approach enables the computation of higher-order moments up to the order of $k$ with machine precision, with a matrix-multiplication cost that is multiplied by a factor of $\frac{(k+1)(k+2)}{2}$ and additional $O(D^6)$ to construct derivatives of the SVD.
\section{Numerical results}\label{sec:results}
In this section, we demonstrate the power of forward AD through numerical calculations for the Ising model using both HOTRG and BWTRG.
\subsection{2D case}
We first consider the two-dimensional classical Ising model on a square lattice.
The Hamiltonian and the partition function are given in terms of the spin variable $\sigma\in\{-1,1\}$:
\begin{equation}
  H(\{\sigma\})=-\sum_{\langle i,j\rangle}\sigma_i\sigma_j,
  \qquad
  Z(\beta)=\sum_{\{\sigma\}}e^{-\beta H(\{\sigma\})}.
  \label{eq:Ising_H_Z}
\end{equation}
where $\langle i,j\rangle$ denotes nearest-neighbor interaction, and $\{\sigma\}$ denotes the sum over all possible spin configurations. $\beta$ denotes the inverse temperature $1/T$.
In the tensor network representation in Eq.~\eqref{eq:tensor_Z}, the local tensor $T$ is explicitly given by
\begin{equation}
  T_{x y x' y'}
  =
  2(\cosh\beta)^2
  (\sqrt{\tanh\beta})^{x+y+x'+y'}
  \,
  \delta_{(x+y+x'+y')\bmod 2,0},
  \qquad
  \label{eq:initial_tensor}
\end{equation}
where $x,y,x',y'\in\{0,1\}$ are bond indices corresponding to the positive and negative directions
of the $x$ and $y$ axes, respectively.
The derivatives of the initial tensor are obtained by analytically differentiating
Eq.~\eqref{eq:initial_tensor} with respect to $\beta$.
Alternatively, one may employ AD libraries to compute the derivatives in the general case. Once we obtain derivatives of initial tensor, we can iteratively apply the forward-mode AD rules.

\begin{figure}[tbp]
  \centering
  \begin{subfigure}[b]{0.48\textwidth}
    \centering
    \includegraphics[width=\textwidth]{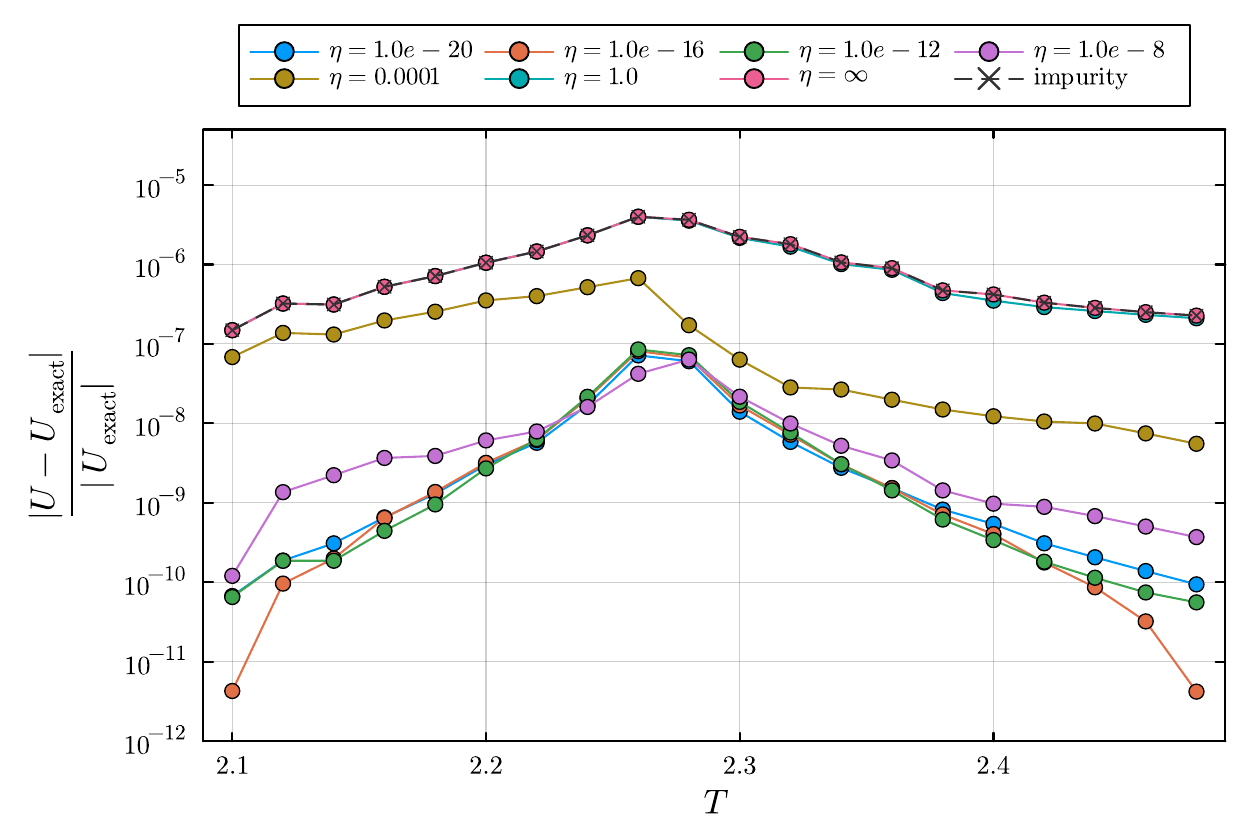}
    \caption{}
    \label{fig:rel_internal_energy}
  \end{subfigure}
  \hfill
  \begin{subfigure}[b]{0.48\textwidth}
    \centering
    \includegraphics[width=\textwidth]{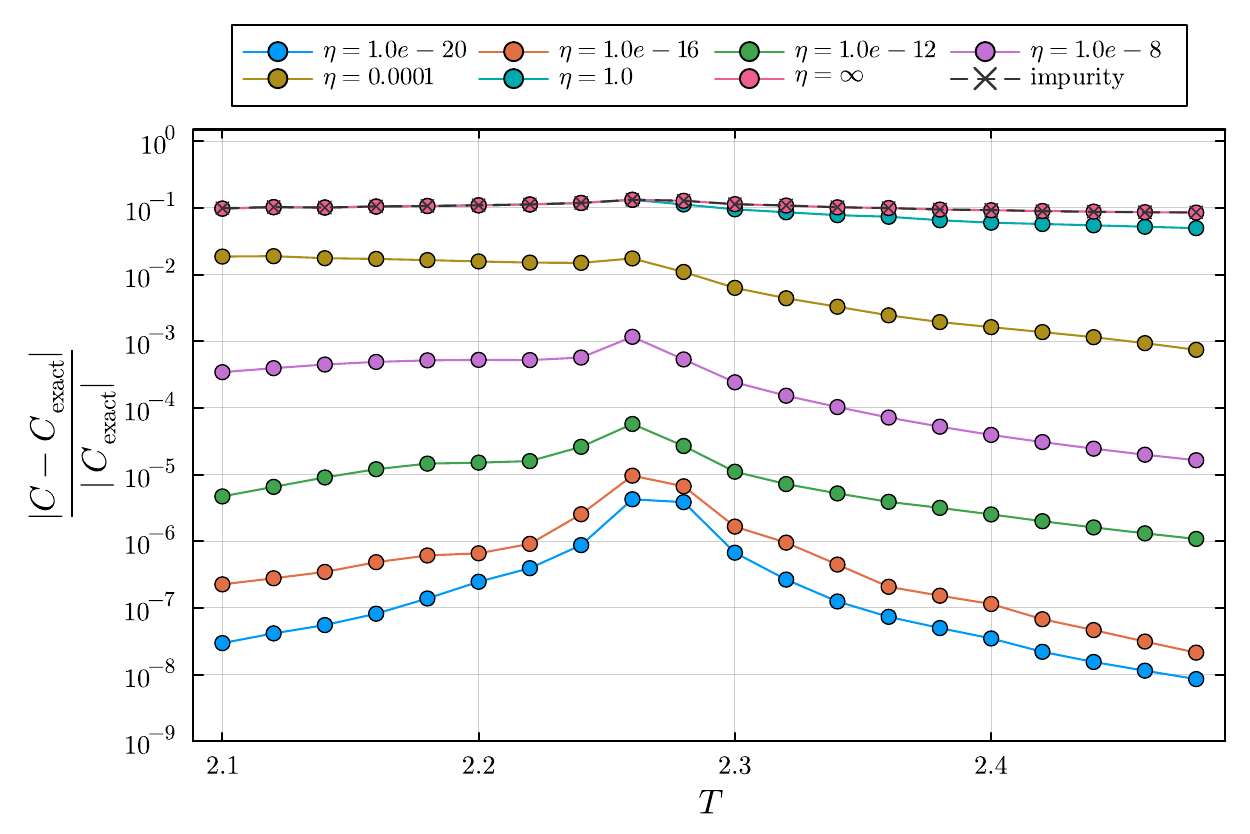}
    \caption{}
    \label{fig:rel_specific_heat}
  \end{subfigure}
\caption{Relative errors of (a) the internal energy and (b) the specific heat obtained from second-order forward-mode AD within HOTRG at $D=80$.
The horizontal axis shows the temperature.
Results are shown for several regularization parameters $\eta$ and are compared with the impurity method.}
  \label{fig:U_and_C}
\end{figure}

We now apply the forward-mode AD to the HOTRG method.
We perform the second-order forward-mode AD for the 2D Ising model using HOTRG with bond dimension $D=80$.
We vary the regularization parameter as
\[
  \eta \in \left\{10^{-20},\,10^{-16},\,10^{-12},\,10^{-8},\,10^{-4},\,10^{0},\,\infty\right\},
\]
and compare the results with those obtained by the impurity method.
We compute the internal energy $U$ and the specific heat $C$ as functions of the temperature $T=1/\beta$.
Figures~\ref{fig:rel_internal_energy} and \ref{fig:rel_specific_heat} show the relative errors of the internal energy $\delta U=\frac{|U-U_{\mathrm{exact}}|}{|U_{\mathrm{exact}}|}$ and the specific heat $\delta C=\frac{|C-C_{\mathrm{exact}}|}{|C_{\mathrm{exact}}|}$ at $V=2^{40}$ with respect to the known exact solutions.
In both figures, the forward-mode AD results are shown by colored circles for each value of $\eta$,
while the impurity results are shown by black symbols.

In Fig.~\ref{fig:rel_internal_energy}, the forward-mode AD results are more accurate than the impurity results.
As $\eta$ decreases, the relative error decreases and approaches the exact value and $\eta\leq 10^{-12}$ provide almost comparable results with each other. For $T=2.1$, the forward-mode AD provides $\delta U \approx O(10^{-11})$ which is almost consistent with exact results.
It is also shown that, as $\eta$ increases, the forward-mode AD results approach those of the impurity method,
and they are consistent in the limit $\eta=\infty$.
This provides numerical evidence that in the HOTRG, the impurity method corresponds to the limit of the forward-mode AD
in which the derivatives of the SVD are effectively set to zero.

The difference becomes more pronounced for the second derivative, i.e., the specific heat, as shown in Fig.~\ref{fig:rel_specific_heat}.
As $\eta$ decreases, the accuracy improves significantly, and the worst performance is again observed for the impurity method.
In particular, the impurity method (equivalently $\eta=\infty$) exhibits about $\mathcal{O}(10^{-1})$ relative errors over a wide temperature range,
whereas the forward-mode AD with $\eta=10^{-20}$ achieves errors smaller than $10^{-5}$.
The improvement is especially clear in both the high- and low-temperature regimes,
where the forward-mode AD calculation is approximately $10^{7}$ times more accurate than the impurity method
at $T=2.248$ and $\eta=10^{-20}$.
These results demonstrate that the forward-mode AD enables substantially higher accuracy than the impurity method
for calculating thermodynamic quantities within the HOTRG.

\begin{figure}[tbp]
  \centering
  \begin{subfigure}[b]{0.48\textwidth}
    \centering
    \includegraphics[width=\textwidth]{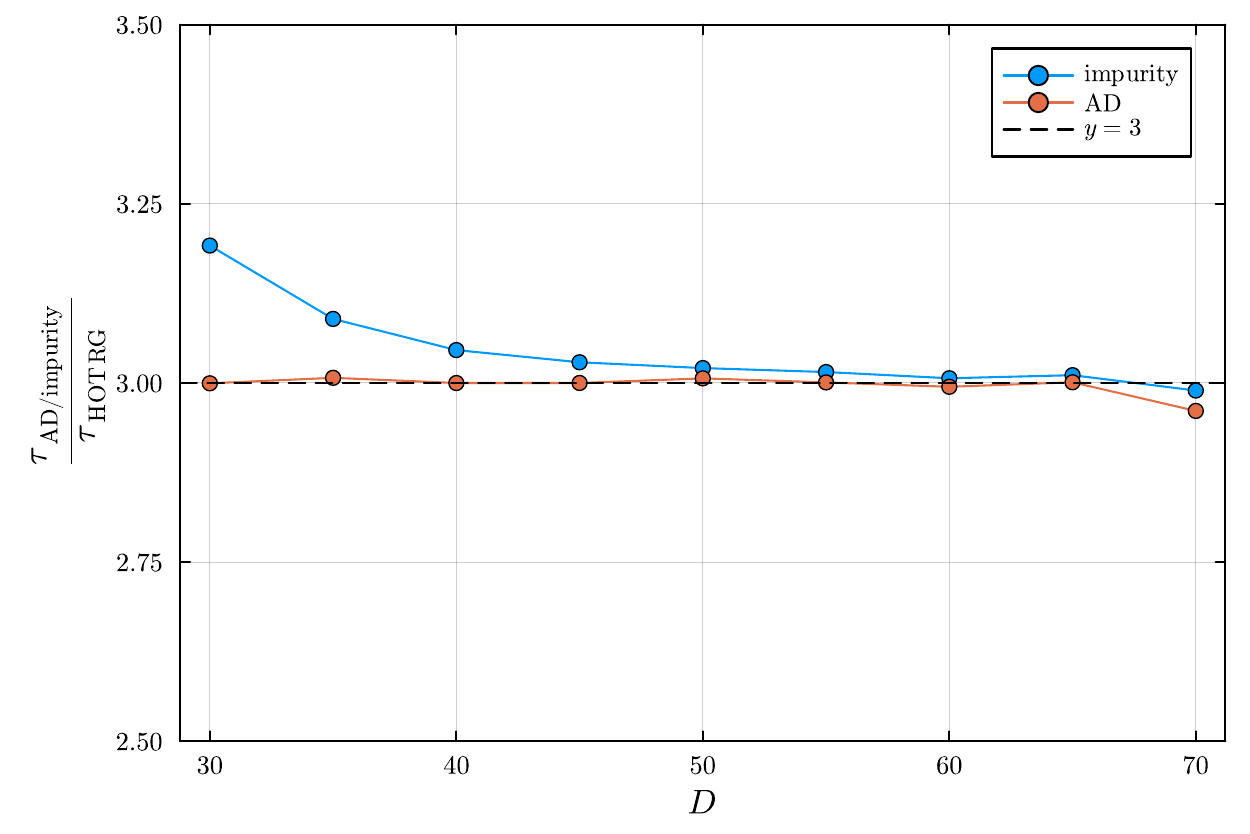}
    \caption{}
    \label{fig:elapsed_cont1}
  \end{subfigure}
  \hfill
  \begin{subfigure}[b]{0.48\textwidth}
    \centering
    \includegraphics[width=\textwidth]{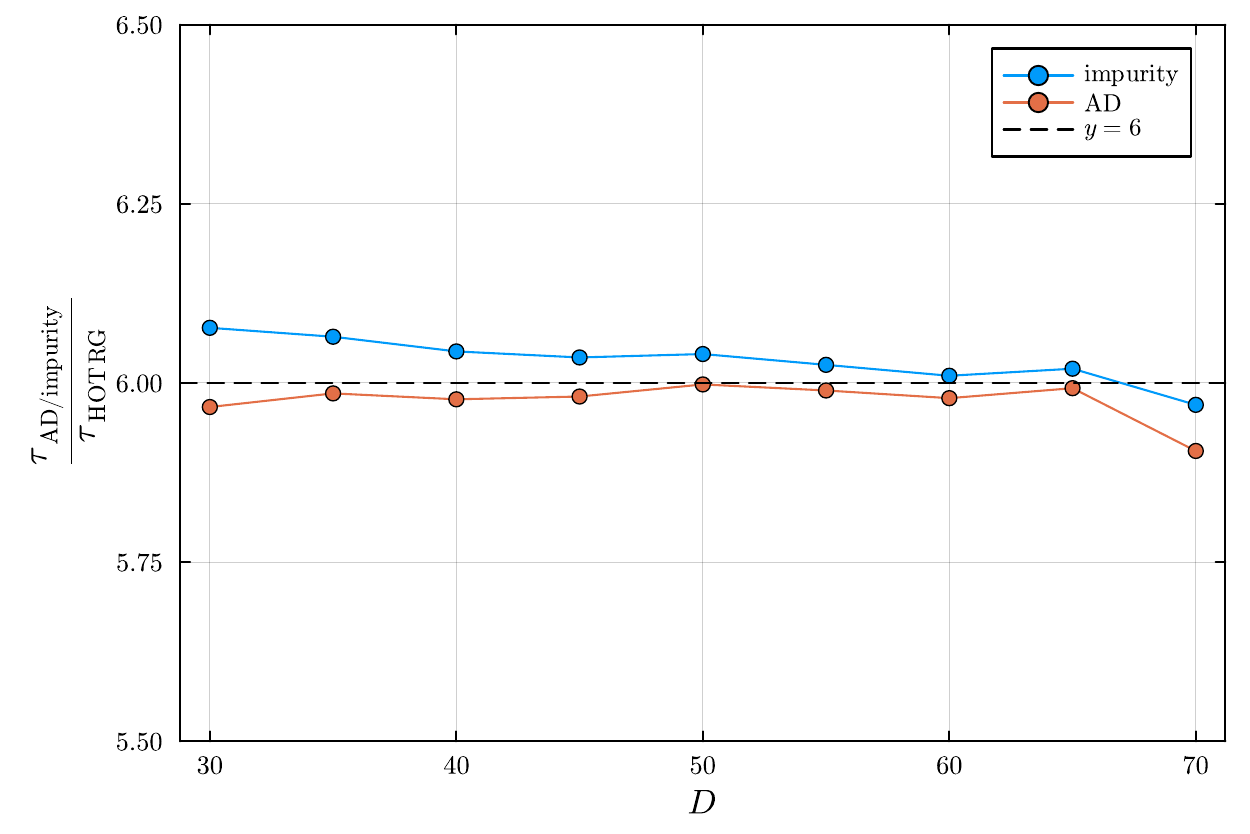}
    \caption{}
    \label{ffig:elapsed_cont2}
  \end{subfigure}
  \caption{Comparison of the elapsed time of the coarse-graining part of the forward-mode AD and the impurity-tensor method as a function of $D$. $\tau_{\mathrm{AD/impurity}}$ denotes the computational time of the AD and impurity method, respectively. $\tau_{\mathrm{HOTRG}}$ denotes the computational time of the HOTRG method alone.
    (a) Elapsed time of the bottleneck contraction part for obtaining up to the first-order derivative, normalized by the original HOTRG contraction time without derivatives.
    (b) Total elapsed time of HOTRG computations up to the second derivatives,
    normalized by the execution time of the original HOTRG.
  The dotted line indicates theoretical scaling factor $\frac{(k+1)(k+2)}{2}$.}
  \label{fig:elapsed_bottleneck}
\end{figure}

\begin{figure}[tbp]
  \centering
  \begin{subfigure}[b]{0.48\textwidth}
    \centering
    \includegraphics[width=\textwidth]{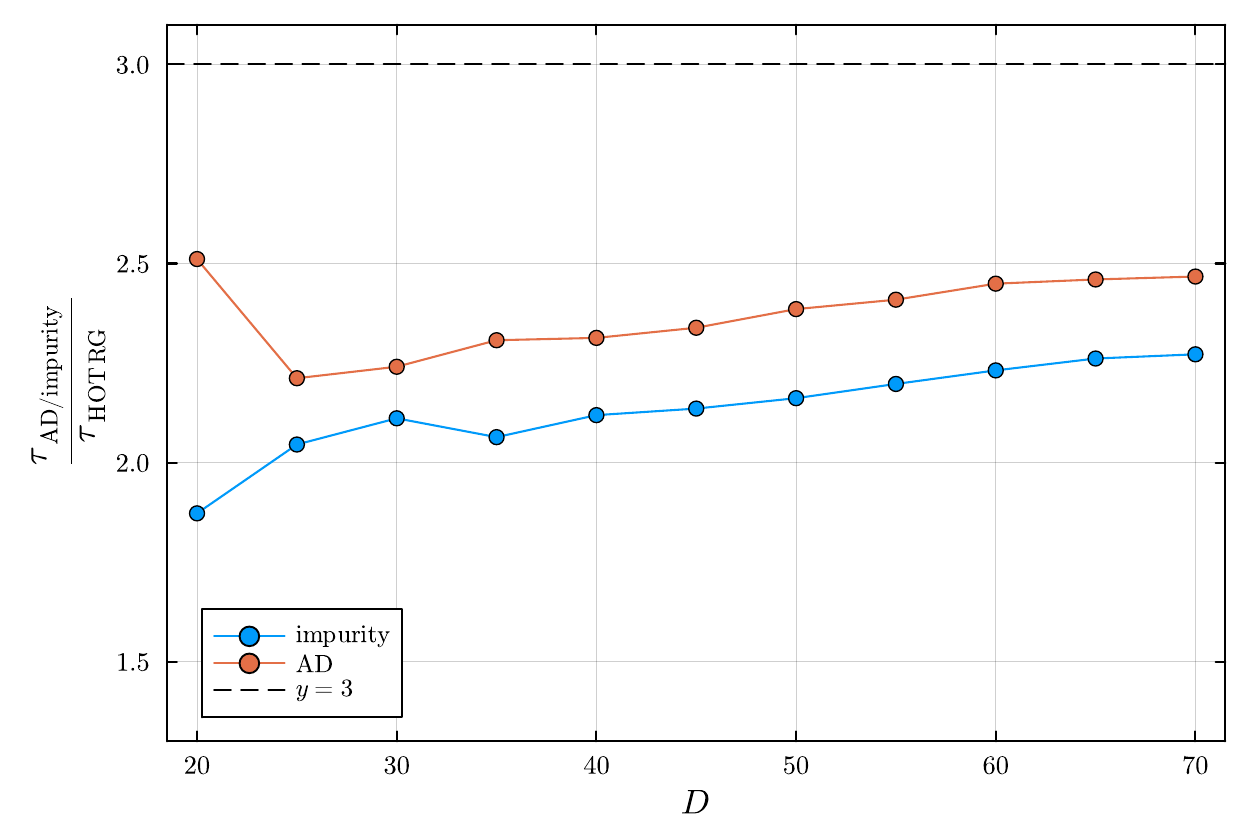}
    \caption{}
    \label{fig:elapsed_bottleneck_1st2nd}
  \end{subfigure}
  \hfill
  \begin{subfigure}[b]{0.48\textwidth}
    \centering
    \includegraphics[width=\textwidth]{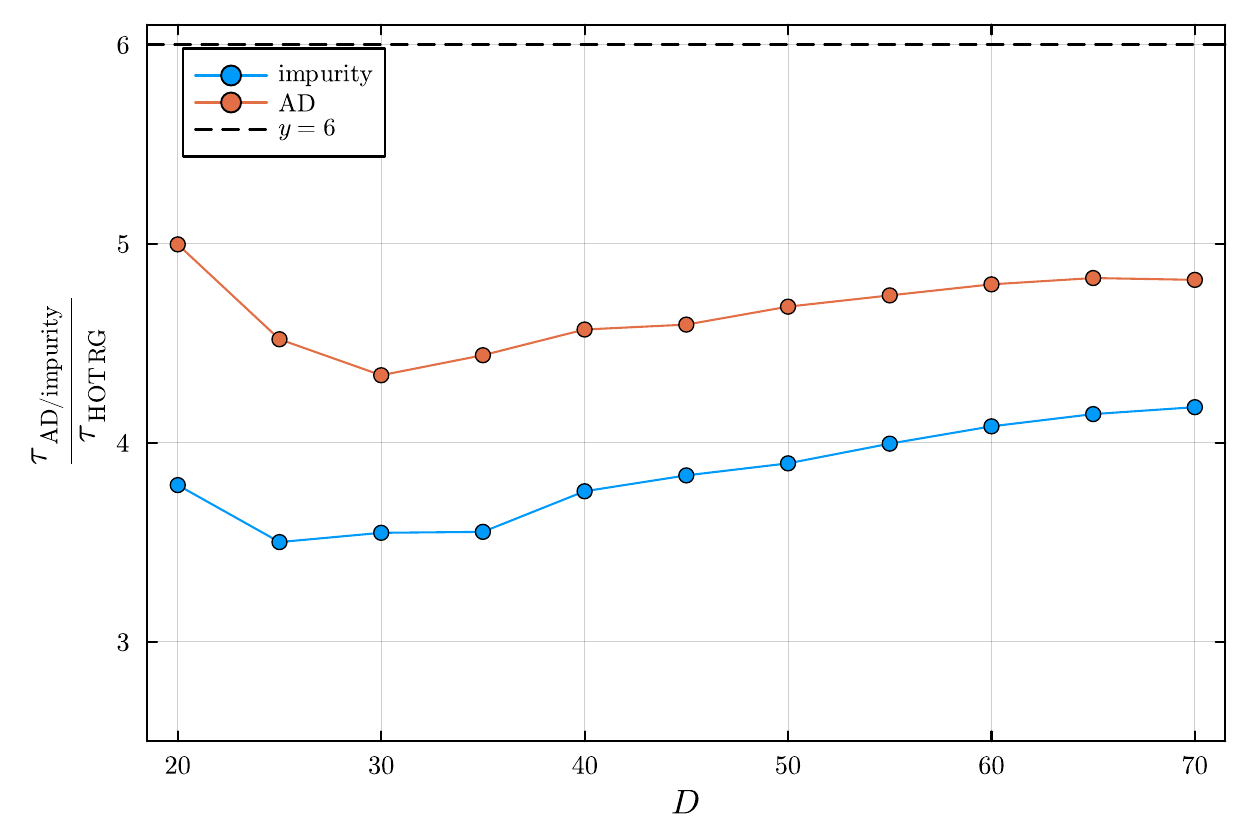}
    \caption{}
    \label{fig:elapsed_HOTRG_total}
  \end{subfigure}
  \caption{Elapsed time comparison between the forward-mode AD and the impurity-tensor method of the HOTRG method as a function of $D$. $\tau_{\mathrm{AD/impurity}}$ denotes the computational time of the AD and impurity method, respectively. $\tau_{\mathrm{HOTRG}}$ denotes the computational time of the usual HOTRG method alone.
    (a) Elapsed time of the obtaining up to first derivatives on a $V=2^{40}$ lattice,
    normalized by the original HOTRG contraction time without derivatives.
    (b) Total elapsed time of HOTRG computations up to the second derivatives on a $V=2^{40}$ lattice,
    normalized by the execution time of the original HOTRG.
  The dotted line indicates theoretical scaling factor $\frac{(k+1)(k+2)}{2}$.}
  \label{fig:elapsed_total}
\end{figure}
Next, we discuss the computational cost.
We first measure the elapsed time of the bottleneck contraction part in Eq.~\eqref{eq:HOTRG} and its derivatives, or impurity update rule.
Figure~\ref{fig:elapsed_bottleneck} shows the elapsed time of the bottleneck contraction in HOTRG for the forward-mode AD computations up to first and second order, and for the corresponding first- and second-order impurity-tensor computations.
These correspond to Eqs.~\eqref{eq:HOTRG}, \eqref{eq:chain rule_HOTRG}, and \eqref{eq:chain rule2_HOTRG} for the forward-mode AD,
and to Eqs.~\eqref{eq:HOTRG}, \eqref{eq:first_imp_rule}, and \eqref{eq:second_imp_rule} for the impurity method. 
For the forward-mode AD, we employ contraction-tree-based differentiation described in Appendix~\ref{ap:general_case}.
All benchmarks are performed on a single CPU using randomly generated dense tensors, and the bond dimension is varied from $D=20$ to $70$.
The vertical axis shows the elapsed time normalized by that of the original HOTRG contraction in Eq.~\eqref{eq:HOTRG}.
The dotted line indicates theoretical scaling factor $\frac{(k+1)(k+2)}{2}$ for derivatives up to the order of $k$.
We find that both the forward-mode AD method and the impurity-tensor method closely follow the expected scaling, indicating that their bottleneck costs are essentially identical, as expected theoretically.

Figure~\ref{fig:elapsed_total} shows the total elapsed time of HOTRG computations up to the first and second derivatives at volume $V=2^{40}$ with $D=20$ to $70$, measured on a single CPU.
The vertical axis shows the elapsed time of each method normalized by the execution time of the original HOTRG without derivatives.
The dotted line again represents theoretical bottleneck scaling factor $\frac{(k+1)(k+2)}{2}$.
In the two-dimensional HOTRG with the squeezer formulation, each coarse-graining step involves three SVDs, and, hence, the subleading costs are non-negligible compared to the bottleneck $O(D^7)$ cost.
As a result, the observed normalized runtime is smaller than the ideal bottleneck scaling.
For both first- and second-derivative cases, the forward-mode AD method becomes slightly more expensive in practice, but it still provides comparable results.
These deviations arise from the additional overhead of computing the squeezers and their derivatives.

Overall, the forward-mode AD exhibits comparable computational scaling and, for the bottleneck contraction, incurs essentially the same computational cost as the impurity method, while achieving significantly higher accuracy.

\begin{figure}[htbp]
  \centering
  \includegraphics[width=0.65\linewidth]{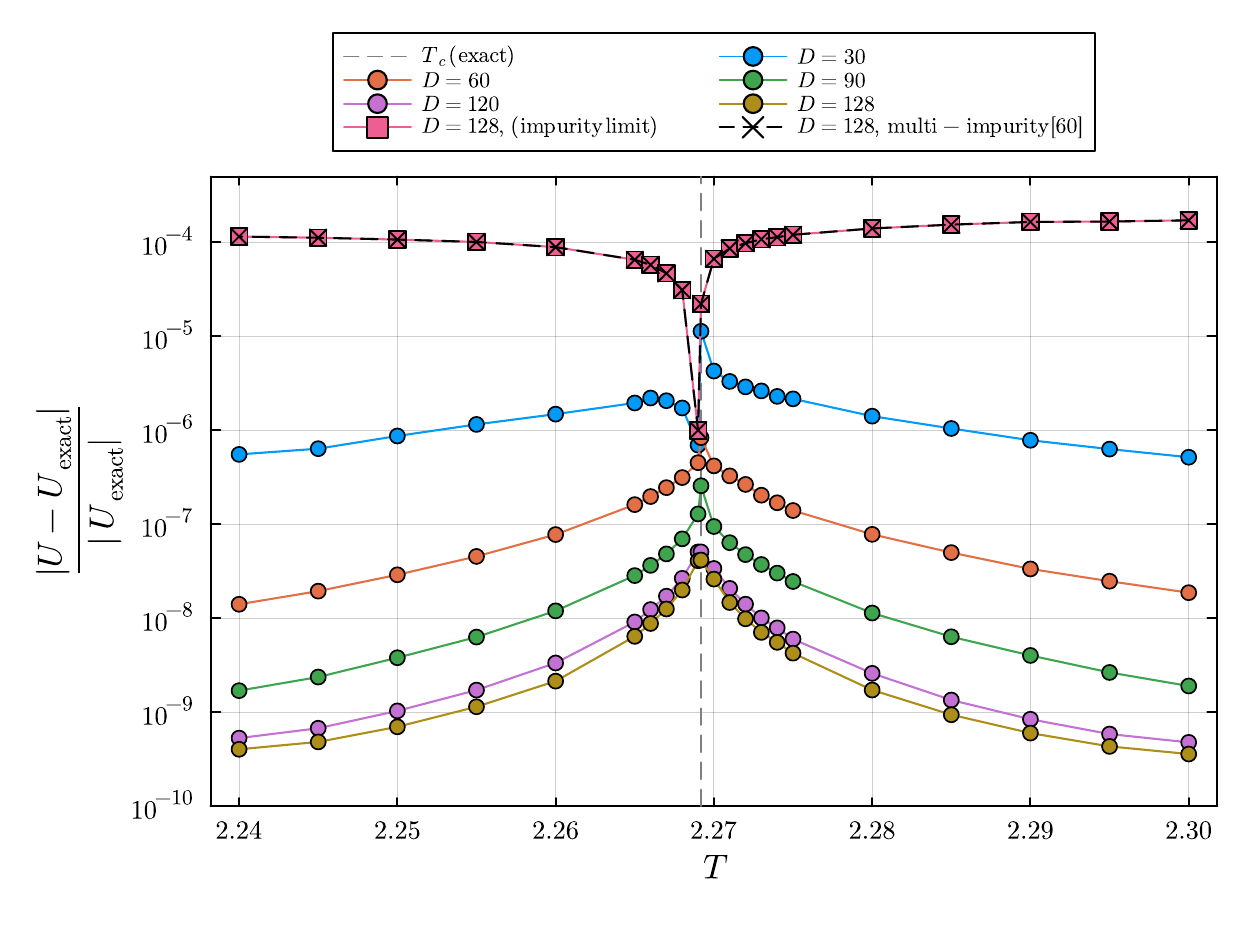}
  \caption{Relative error of the internal energy with respect to the exact result as a function of temperature.
  Circles show the forward-mode AD data for various bond dimensions.
  The impurity limit corresponds to setting the SVD derivatives to zero and using the bond-based initial tensor described in Sec.~\ref{subsec:BWTRG}; these data are shown by square symbols.
 The multi-impurity data from Refs.~\cite{moritaMultiimpurityMethodBondweighted2024,morita_github} are converted into relative errors and shown as black symbols. The dotted line indicates the exact critical temperature.}
  \label{fig:BWTRG_U}
\end{figure}

Next, we investigate the BWTRG case. Figure~\ref{fig:BWTRG_U} shows the relative error of the internal energy with respect to the exact value,
computed using the forward-mode AD formulation of BWTRG with $\eta=10^{-20}$.
The bond dimension is varied from $D=30$ to $128$.
For comparison, we also include the $D=128$ multi-impurity BWTRG data reported in Ref.~\cite{moritaMultiimpurityMethodBondweighted2024}.
We further present the impurity limit of the forward-mode AD formulation, in which we use an alternative representation of the initial tensor and set the derivatives of the SVD to zero, as described in Sec.~\ref{subsec:BWTRG}. 
Even at $D=30$, the forward-mode AD results are more accurate than the multi-impurity result at $D=128$, and they systematically improve as $D$ increases. In the impurity limit, our results are consistent with the multi-impurity results obtained with the triad representation of BWTRG.
We note that our BWTRG implementation is based on the original Levin--Nave-type TRG~\cite{levinTensorRenormalizationGroup2007,adachiBondweightedTensorRenormalization2022} and does not use a partial SVD; accordingly, the computational cost scales as $O(D^6)$. In contrast, Ref.~\cite{moritaMultiimpurityMethodBondweighted2024} employs a triad representation, leading to an $O(D^5)$ scaling. Nevertheless, we have confirmed that the two implementations yield consistent free energies at $D=128$. We also confirmed that the two representations of the initial tensor yield nearly identical free energies and internal energies for finite $\eta$. 
Another remark is that the impurity limit based on the initial tensor in Eq.~\eqref{eq:initial_tensor} does not correspond to a systematic summation of impurity-tensor networks, and, therefore, we could not find meaningful results.
These observations indicate that, when there exists a correspondence between the impurity method and the limit in which the SVD derivatives are set to zero, the impurity method serves as a theoretical lower bound on the accuracy of the forward-mode AD formulation.
Finally, we note that the original BWTRG study~\cite{adachiBondweightedTensorRenormalization2022} already used automatic differentiation to compute the internal energy. Here, we include this comparison to demonstrate that our explicit forward-mode formulation reproduces accurate results.
\subsection{Some applications}
Since TRG based on the forward-mode AD directly calculates derivatives of renormalized tensor at each coarse-graining step, it is also possible to differentiate quantities obtained from renormalized tensors, not only the derivatives of the partition function.
The Gu--Wen ratio or partition function ratio $X$~\cite{guTensorEntanglementFilteringRenormalizationApproach2009} is an important quantity computed from the renormalized tensor. The definition of $X$ at $n$th coarse-graining step is\footnote{Here we denote the renormalized tensor $T^{(n)}$ including bond weights.}
\begin{equation}\label{eq:X}
  X^{(n)}=\frac{\left(\vcenter{\hbox{\includegraphics[scale=0.7]{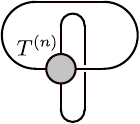}}}\right)^2}{\vcenter{\hbox{\includegraphics[scale=0.7]{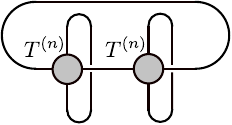}}}}.
\end{equation}
It effectively measures the ground-state degeneracy and, therefore, serves as an order parameter for the Ising model.
It was reported that the universal value of $X$ at the critical point is determined by the conformal field theory~\cite{moritaMultiimpurityMethodBondweighted2024,moritaTensorRenormalizationGroup2025}.
It has also been reported that partition function ratio can be extended to systems with continuous symmetries~\cite{akiyamaTensorRenormalizationGroup2026}.
In Ref.~\cite{moritaMultiimpurityMethodBondweighted2024}, it is shown that, since $X$ is a dimensionless quantity, it obeys the finite-size scaling form
\begin{align}\label{eq:scale_X}
  X = g\left(L^{1/\nu}\tau\right),
\end{align}
with $\tau = (T-T_c)/T_c$, where $T_c$ is the critical temperature, $g$ is a scaling function, and $\nu$ is the critical exponent.
By differentiating Eq.~\eqref{eq:scale_X} with respect to $T$, we obtain
\begin{align}\label{eq:scale_dX}
  \pdv{X}{T}
  = \frac{1}{T_c} L^{1/\nu} g'\left(L^{1/\nu}\tau\right).
\end{align}
From Eq.~\eqref{eq:scale_dX}, at $T=T_c$ ($\tau=0$) we have
\begin{align}\label{eq:log_scale_dX}
  \log\left|\pdv{X}{T}\right|_{T=T_c}
  = \frac{1}{\nu}\log L + A,
\end{align}
where $A$ is constant. Eq.~\eqref{eq:log_scale_dX} predicts a linear dependence of $\log\left|\pdv{X}{T}\right|_{T=T_c}$ on $\log{L}$ with slope of $1/\nu$. 
Therefore, by evaluating Eq.~\eqref{eq:log_scale_dX} for multiple system sizes at critical temperature, we can extract $1/\nu$ from a linear fit.
A similar analysis was used for a $q$-state Potts model by impurity method with the Binder parameter~\cite{moritaCalculationHigherorderMoments2019}. 
The diagrammatic representation of $\pdv{X}{T}$ at $n$th coarse-graining step is given by
\begin{equation}
  \pdv{X^{(n)}}{T}=-2\beta^2\left[\,\frac{\vcenter{\hbox{\includegraphics[scale=0.7]{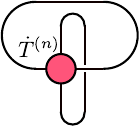}}}\,\,\vcenter{\hbox{\includegraphics[scale=0.7]{fig_old/X_n.pdf}}}}{\vcenter{\hbox{\includegraphics[scale=0.7]{fig_old/X_d.pdf}}}}-\frac{\vcenter{\hbox{\includegraphics[scale=0.7]{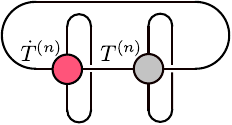}}}\left(\vcenter{\hbox{\includegraphics[scale=0.7]{fig_old/X_n.pdf}}}\right)^2}{\left(\vcenter{\hbox{\includegraphics[scale=0.7]{fig_old/X_d.pdf}}}\right)^2}\,\right].
\end{equation}
Since our forward-mode AD formulation can calculate derivatives of $T$ at every TRG step, a calculation at the critical temperature directly gives the volume dependence of $\pdv{X^{(n)}}{T}$. We compute the critical exponent of the Ising model by a linear fit of Eq.~\eqref{eq:log_scale_dX}. 
The critical temperature is determined as the point at which $X$ becomes discontinuous, or where it takes an intermediate value that is approximately volume independent at $L=2^{30}$.

\begin{figure}[tbp]
  \centering
  \begin{subfigure}[t]{0.49\linewidth}
    \centering
    \includegraphics[width=\linewidth]{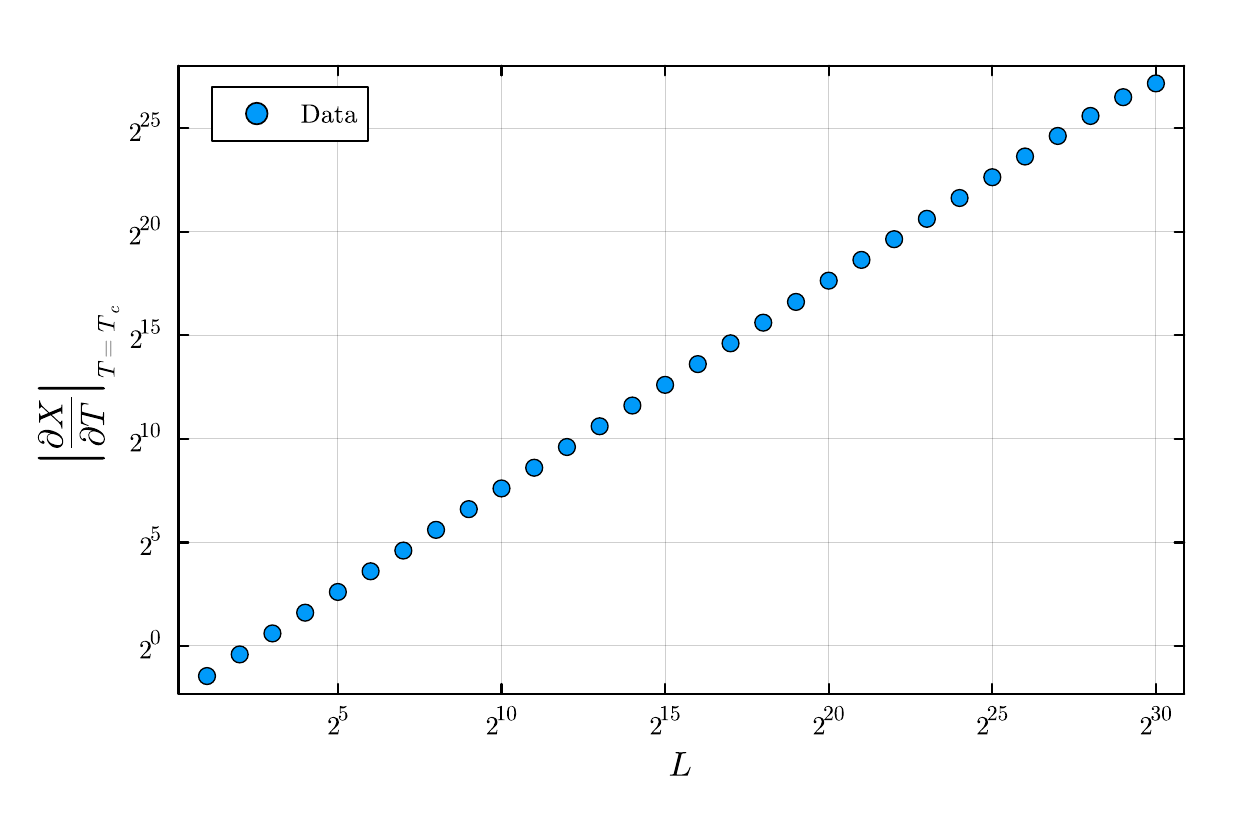}
    \caption{}
    \label{fig:dX_vs_L}
  \end{subfigure}\hfill
  \begin{subfigure}[t]{0.49\linewidth}
    \centering
    \includegraphics[width=\linewidth]{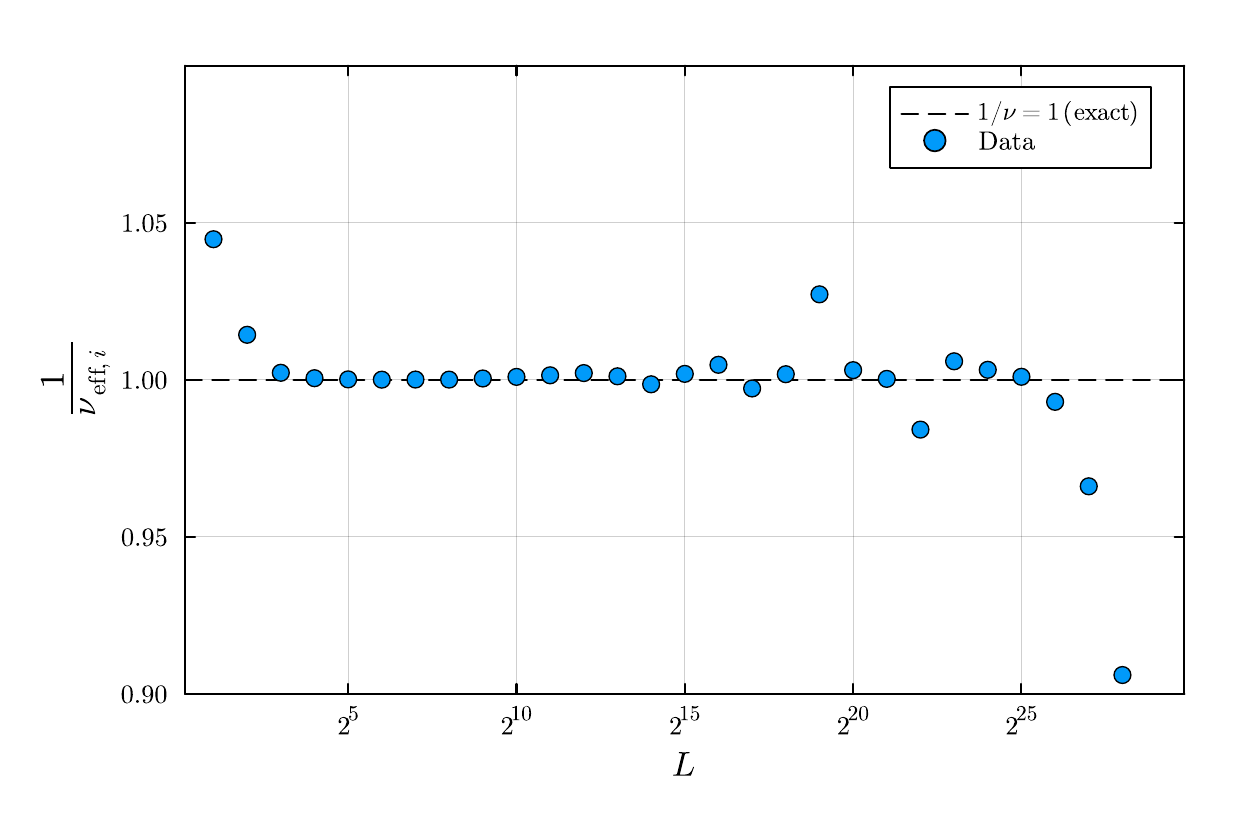}
    \caption{ }
    \label{fig:bloc_vs_L}
  \end{subfigure}
\caption{Two-dimensional Ising model computed using the forward-mode AD BWTRG at $D=130$.
(a) $L$ dependence of $\left|\pdv{X}{T}\right|_{T=T_c}$.
(b) $L$ dependence of $1/\nu_{\mathrm{eff},i}$.}
\end{figure}

\begin{figure}[tbp]
  \centering
  \includegraphics[width=0.65\linewidth]{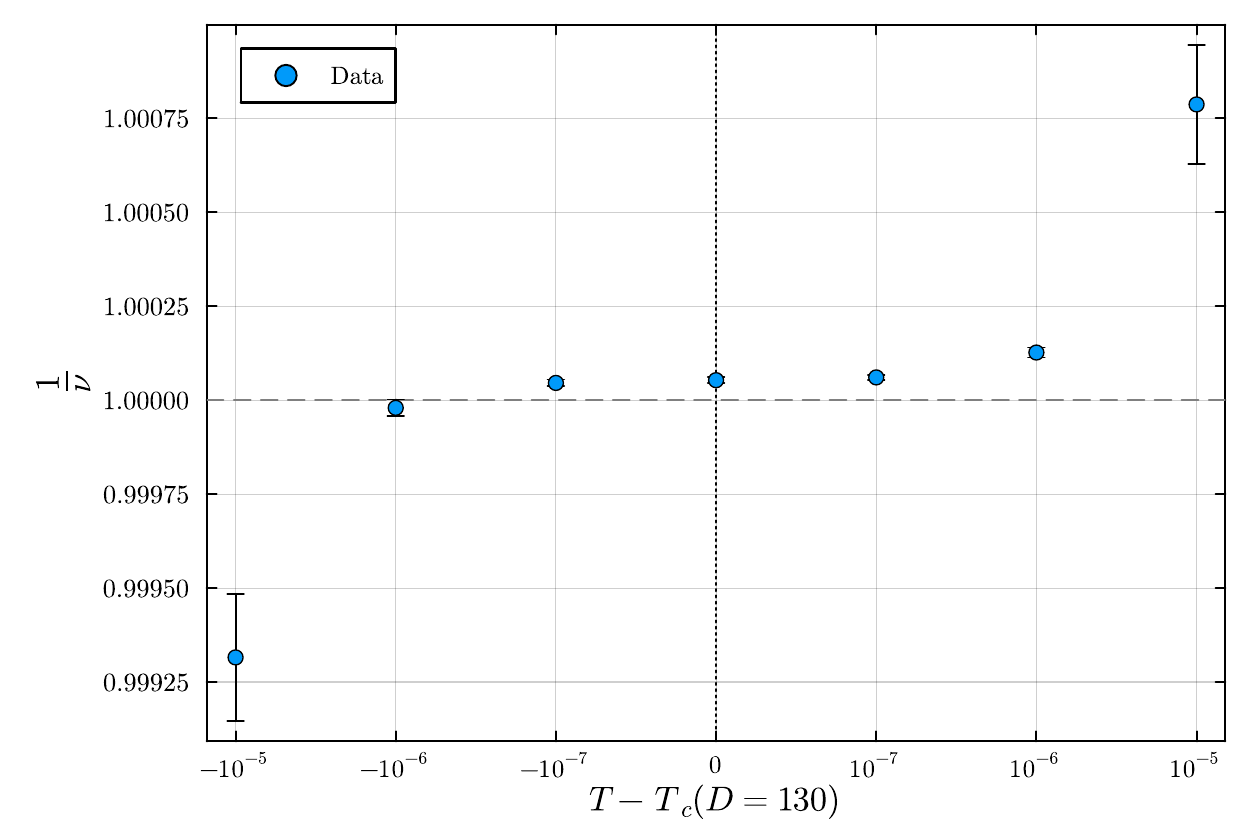}
  \caption{The critical exponent $1/\nu$ at $D=130$ obtained using the forward-mode AD BWTRG, plotted as a function of $T - T_c$, where $T_c(D=130)=2.2691853$.}
  \label{fig:BWTRG_nu_130}
\end{figure}

Figure~\ref{fig:dX_vs_L} shows typical behavior of the data obtained by BWTRG at $D=130$ and the critical temperature $T_c(D=130)$=2.2691853 estimated by $X$. Figure~\ref{fig:bloc_vs_L} shows an effective critical exponent measured by
\begin{equation}\label{eq:effective_nu}
\frac{1}{\nu_{\mathrm{eff},i}}=\frac{\log_2 \left|\pdv{X}{T}\right|_{i+1}-\log_2 \left|\pdv{X}{T}\right|_i}{\log_2 L_{i+1}-\log_2 L_i}
\end{equation}
for all data points. 
The data exhibit a plateau near $1/\nu=1$ for small system sizes.
As the system size increases, the data points fluctuate; this is likely due to finite bond dimension effects~\cite{uedaFinitesizeFiniteBond2023} and the limited resolution in the estimated critical temperature.
Using data from $L=2^5$ to $2^{9}$, we obtain $\frac{1}{\nu}(D=130) = 1.000053(8)$, which is close to the exact value $1/\nu=1$. 
Figure \ref{fig:BWTRG_nu_130} shows the temperature dependence of the estimated critical exponent around $T_c(D=130)$ for $D=130$. The horizontal axis represents $T-T_c$. As the resolution of the estimated critical temperature is improved, the results become more stable with respect to temperature. Even at $T-T_c(D=130)=10^{-5}$, we obtain $1.00079(16)$, which differs from the exact value only by $O(10^{-4})$.

\begin{figure}[tbp]
  \centering
  \includegraphics[width=0.65\linewidth]{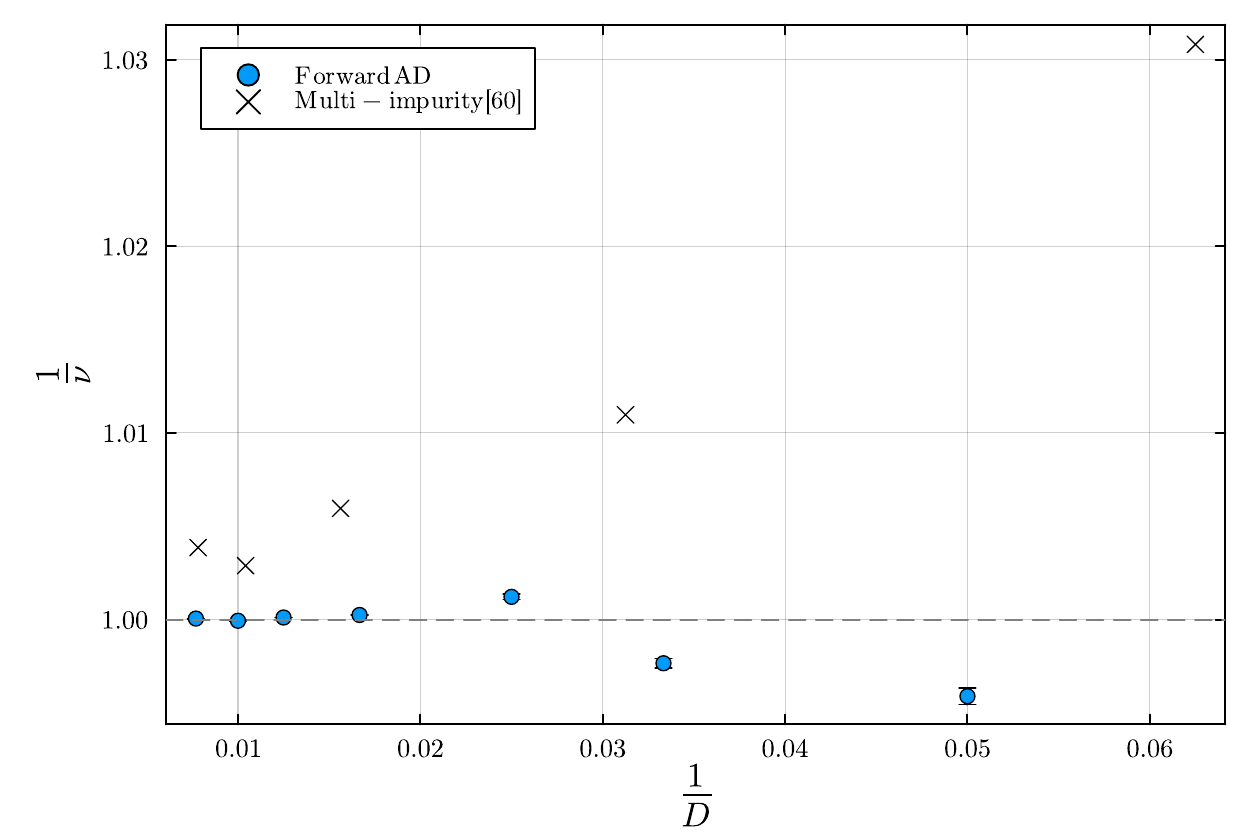}
  \caption{$1/D$ dependence of the critical exponent $1/\nu$ obtained from the forward-mode AD BWTRG (circles).
  For comparison, the results from the scaling-collapse analysis taken from Refs.~\cite{moritaMultiimpurityMethodBondweighted2024,morita_github} are shown by black symbols. The dotted line indicates the exact value.}
  \label{fig:BWTRG_nu}
\end{figure}
We also show the $1/D$ dependence of $1/\nu$ together with the result in Ref.~\cite{moritaMultiimpurityMethodBondweighted2024}, obtained by a scaling-collapse analysis of $X$ around the critical temperature (see Fig.~\ref{fig:BWTRG_nu}). The critical temperature is calculated with a resolution of $5\times 10^{-8}$ for all bond dimensions in order to examine the $D$ dependence precisely.
The estimated $1/\nu$ approaches the exact value asymptotically, and the results are more stable and accurate than those reported in Ref.~\cite{moritaMultiimpurityMethodBondweighted2024}. These results indicate that performing a linear fit to $\partial X/\partial T$ provides a useful and reliable means of extracting the critical scaling behavior.

We also extend the forward-mode AD formulation to the three-dimensional HOTRG.
In the following, we define $X^{(n)}$ as
\begin{equation}
    X^{(n)} = \frac{\left(Z(L_x,L_y,L_z)\right)^2}{Z(2L_x,L_y,L_z)},
\end{equation}
where $Z(L_x,L_y,L_z)$ is the partition function obtained after the $n$th coarse-graining step with $L_x=L_y=L_z=2^n$, and the denominator represents the spatially extended partition function in the $\hat{x}$ direction. We always perform coarse-graining by HOTRG in the order $\hat{x}, \hat{y}, \hat{z}$.

\begin{figure}[tbp]
  \centering
  \begin{subfigure}[t]{0.49\linewidth}
    \centering
    \includegraphics[width=\linewidth]{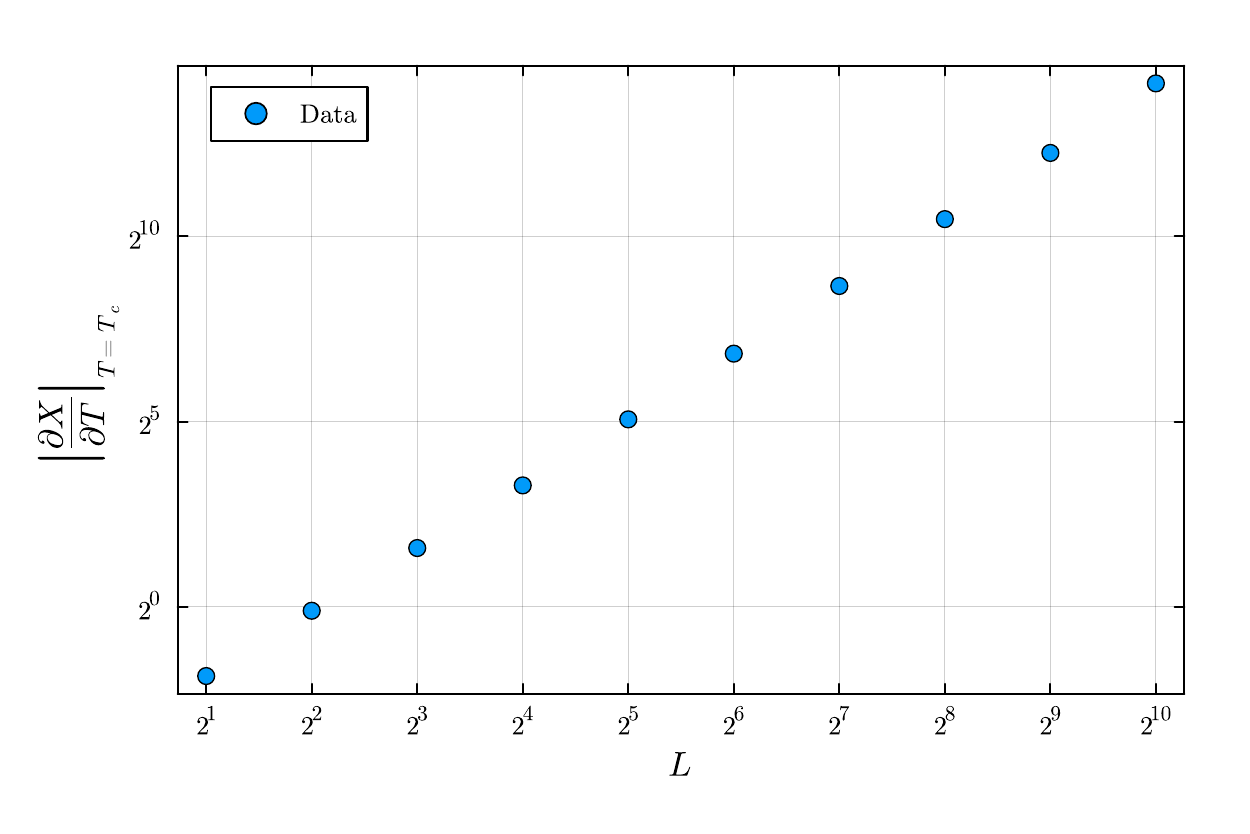}
    \caption{ }
    \label{fig:dX_vs_L_3D}
  \end{subfigure}\hfill
  \begin{subfigure}[t]{0.49\linewidth}
    \centering
    \includegraphics[width=\linewidth]{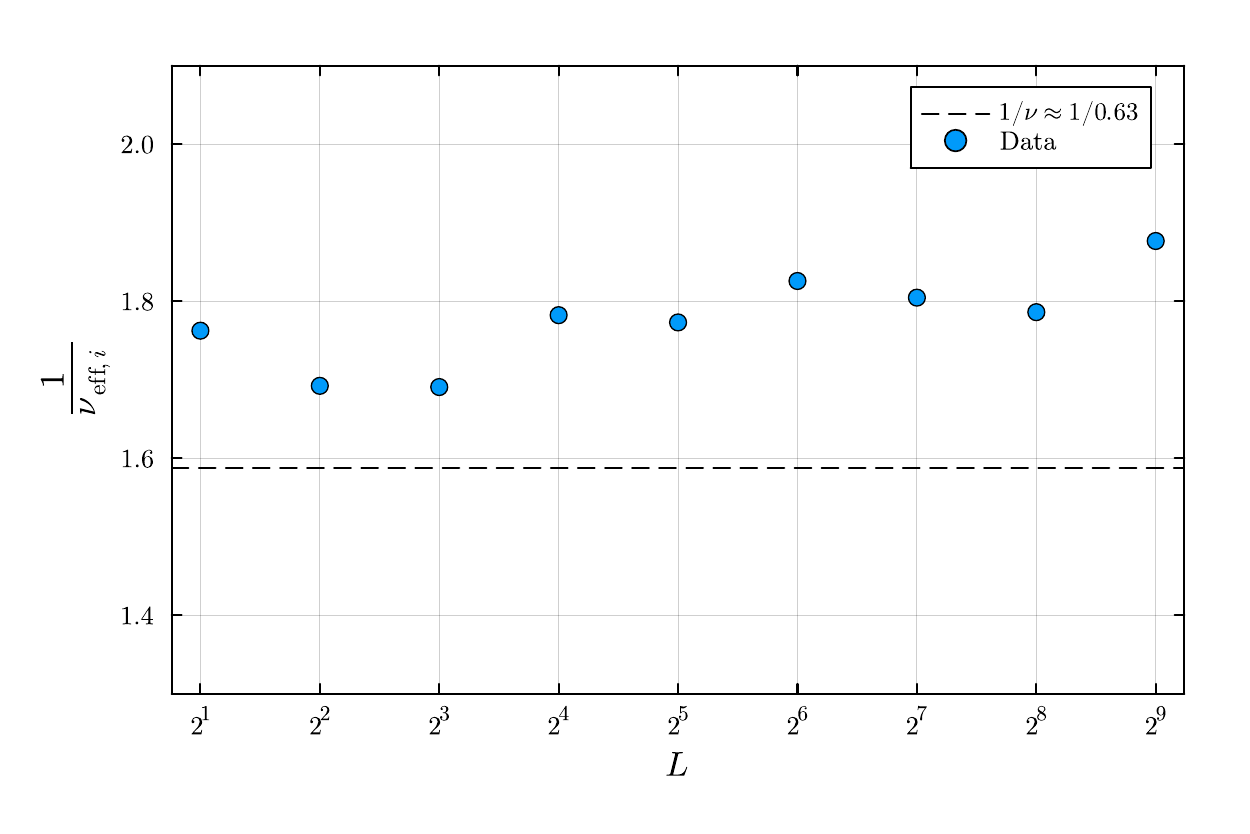}
    \caption{ }
    \label{fig:bloc_vs_L_3D}
  \end{subfigure}
\caption{Three-dimensional Ising model computed using the forward-mode AD HOTRG at $D=32$.
(a) $L$ dependence of $\left|\pdv{X}{T}\right|_{T=T_c}$.
(b) $L$ dependence of $1/\nu_{\mathrm{eff},i}$.}
\end{figure}

We perform calculations for the three-dimensional Ising model at $D=32$ and at the estimated critical temperature $T_c(D=32)=4.5079$, determined from $X$. The volume dependence of $\left|\partial X/\partial T\right|_{T=T_c}$ and $1/\nu_{\mathrm{eff},i}$ is shown in Figs.~\ref{fig:dX_vs_L_3D} and~\ref{fig:bloc_vs_L_3D}. Even in the three-dimensional case, $\log\left|\partial X/\partial T\right|_{T=T_c}$ exhibits linear scaling with respect to $\log L$. However, $1/\nu_{\mathrm{eff},i}$ shows a strong dependence on $L$ and no clear plateau is observed, unlike in the two-dimensional case. From a linear fit for $L=2^2$ to $2^7$, we obtain $\nu(D=32)=0.571(4)$.

\begin{figure}[tbp]
  \centering
  \includegraphics[width=0.65\linewidth]{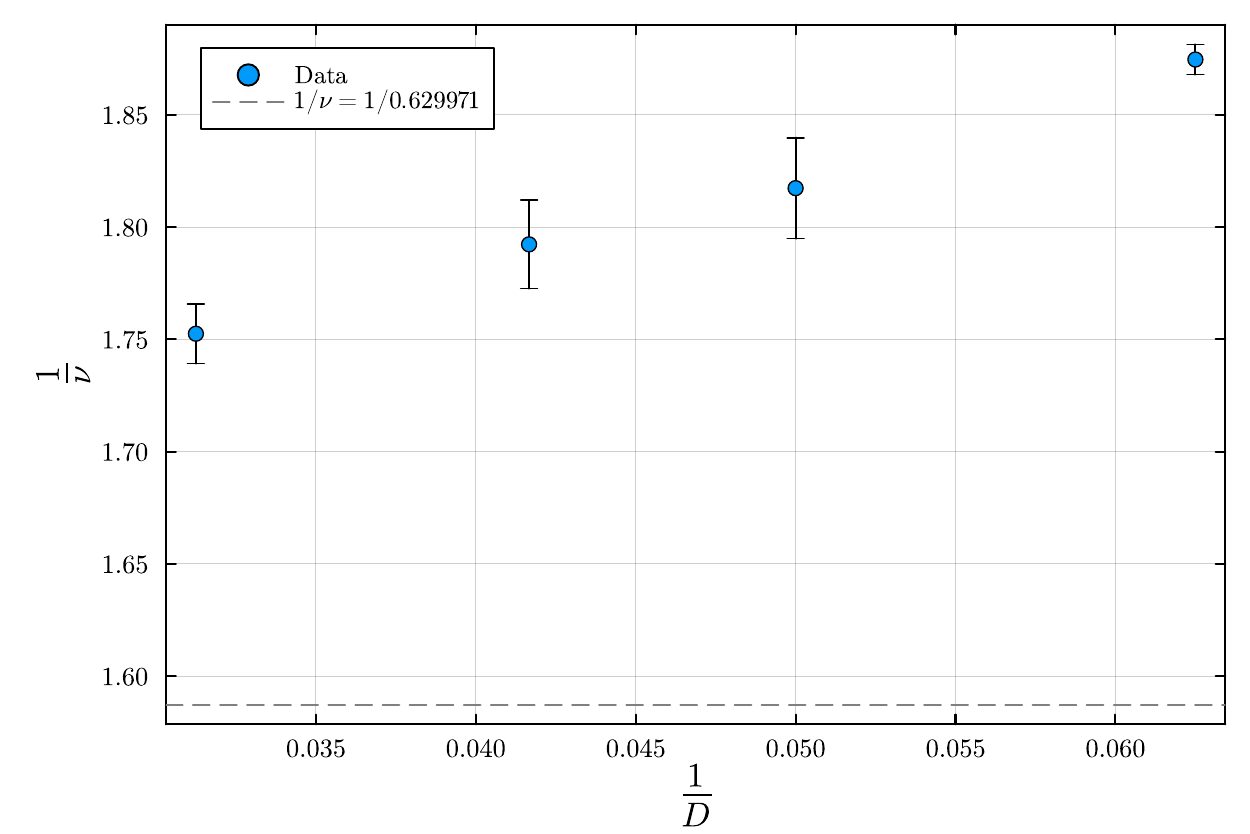}
  \caption{$1/D$ dependence of the critical exponent $1/\nu$ obtained from the 3D HOTRG. The dotted line indicates the known value from Ref.~\cite{Kos:2016ysd}.}
  \label{fig:nu_vs_chi_3d}
\end{figure}

Figure~\ref{fig:nu_vs_chi_3d} shows the $1/D$ dependence of the estimated $1/\nu$ together with the known value $\nu=0.629971(4)$ obtained from the conformal bootstrap~\cite{Kos:2016ysd}. The critical temperatures are estimated with a resolution of $10^{-4}$. As $D$ increases, $1/\nu$ becomes closer to the known value. However, at small bond dimensions, we observed the estimated value strongly depends on the choice of the volume range and on the definition of $X$, namely, on which direction is extended.
We also comment on the resolution of $T_c$. Since a more precise determination of  $T_c $ is computationally expensive, we performed the calculation with a resolution of only $10^{-4}$, which is of the same order as the convergence of $T_c(D)$ with respect to $D$. Therefore, the estimated value also includes a systematic error coming from how close the estimated critical temperature is to the actual $T_c$ at fixed $D$.

From the above discussion, we conclude that, in the three-dimensional case, the estimate is affected by several additional sources of systematic error.
We attribute the deviation from the known value, as well as the instabilities discussed above, to the limited bond dimension in the three-dimensional HOTRG calculation.
We also find that the specific heat is not smooth at accessible $D$ in three dimensions (data not shown), which is likely due to discontinuities in the SVD differentiation and numerical instability associated with the large degeneracy of singular values as well as the limited bond dimension in the three-dimensional Ising model.
We leave further investigation at larger $D$ for future work.

\section{Discussions and summary}\label{sec:summary}
In this paper, we have proposed a simple framework for incorporating automatic differentiation into tensor renormalization group methods based on forward-mode AD.
In our formulation, the memory cost for derivatives up to the order of $k$ scales as $k+1$ times that of the original TRG, independent of the graph depth while the contraction cost increases by a factor $(k+1)(k+2)/2$.
The proposed forward-mode AD framework has been shown to be applicable to various TRG schemes, including HOTRG and BWTRG.
In the limit where derivatives of the SVD are neglected, a theoretical correspondence to impurity methods can be established.
In numerical calculations, derivatives of the partition function, including the internal energy and specific heat, have been accurately evaluated using the forward-mode AD formulation.
In the impurity limit, consistency with established impurity approaches for both HOTRG and BWTRG is confirmed.
Moreover, forward-mode AD achieves significantly higher accuracy than the impurity method, which neglects the parameter dependence of the isometric tensor~\cite{moritaMultiimpurityMethodBondweighted2024}.
We have also shown that, for the bottleneck tensor contraction in each coarse-graining step of the HOTRG,
the computational cost for evaluating derivatives up to the order of $k$ increases by a factor of $(k+1)(k+2)/2$, while the total cost of the HOTRG is slightly higher than that of the impurity method, due to the additional evaluations associated with differentiating the SVD or the squeezers.
In Appendix~\ref{ap:general_case}, we have shown that the same scaling behavior can be achieved for general tensor networks.

The forward-mode AD based TRG enables finite-size scaling analyses of derivatives of the Gu--Wen ratio,
yielding a highly accurate estimate of the critical exponent $1/\nu$.
This is a key advantage of the forward-mode AD, because the volume dependence of physical observables is naturally accessible within a single forward pass of the TRG flow. 

We have also extended our algorithms to three-dimensional cases, where the increasing cost with $D$ limits the accessible bond dimension which leads to a less accurate estimate of $\nu$.

Finally, we discuss degenerate singular values and AD.
In the presence of degenerate singular values at the truncation threshold $D$, the TRG update map can become discontinuous with respect to $\beta$, and its derivatives may, therefore, also become discontinuous.
In tensor network calculations, truncating a degenerate multiplet should be avoided, because the singular value spectrum typically encodes symmetry information of the system~\cite{whiteDensitymatrixAlgorithmsQuantum1993b}. Cutting a degenerate multiplet midway explicitly breaks the symmetry and may drive the renormalization flow toward an incorrect fixed-point tensor.
Moreover, degeneracy implies a rotational freedom within the degenerate subspace: The singular vectors are not uniquely defined, but are determined only up to a unitary transformation. This internal freedom can be viewed as a gauge freedom. If one truncates inside the degenerate subspace, the resulting truncation projector becomes gauge dependent, leading to numerical instabilities and inconsistencies among different implementations or algorithms.
Although one may attempt to choose $D$ so as to avoid truncating degenerate singular values, it is, in general, difficult to guarantee this condition at every TRG step. A possible remedy is to enforce gauge fixing by aligning the singular vectors continuously as $\beta$ is varied, which can improve numerical stability and apparent smoothness in $\beta$~\cite{blochTensorRenormalizationGroup2021,francuzStableEfficientDifferentiation2025}. However, such a procedure may still break the symmetry encoded in the tensor if the truncation itself splits the degenerate multiplet. Another strategy is to keep or discard the entire degenerate sector at the cutoff, effectively allowing the bond dimension to vary. While this preserves the symmetry, it introduces a $\beta$-dependent effective bond dimension and can, therefore, lead to discontinuities in the truncation scheme. In our 3D HOTRG calculations, this issue is more severe because the accessible bond dimension $D$ is not large enough to suppress such truncation-induced discontinuities, resulting in discontinuous specific heat.
A similar issue arises when evaluating order parameters associated with $Z_2$ symmetry in the Ising model, such as the squared magnetization.
Introducing an external source $h$ in the Hamiltonian allows one, in principle, to apply AD and then take the limit $h\to 0$.
In practice, however, we found that the squared magnetization at exactly $h=0$ is unstable with respect to the RG steps due to the instability of the SVD derivatives associated with degenerate singular values.
This instability can be mitigated by introducing an extremely small but finite $h$, at the cost of explicitly breaking the spin-flip symmetry.
In the impurity limit, the instability is avoided because derivatives associated with the SVD are entirely neglected, however, this does not necessarily imply improved accuracy.

A natural next step is to push to larger bond dimensions in three or four dimensions by incorporating symmetry-blocking techniques~\cite{devosTensorKitjlJuliaPackage2025},
or by adopting linear renormalization schemes~\cite{uedaGlobalTensorNetwork2025}.
We also plan to extend the method to magnetization or more generally, correlation functions in a stable way~\cite{francuzStableEfficientDifferentiation2025}.
It is also interesting to extend the forward-mode AD to optimization-based TRG algorithms, for example, TRG with randomized SVD~\cite{moritaTensorRenormalizationGroup2018,nakayamaRandomizedHigherorderTensor2023}, which is particularly useful in higher dimensions, and Tensor Network Renormalization-type algorithms~\cite{evenblyTensorNetworkRenormalization2015,yangLoopOptimizationTensor2017,hauruRenormalizationTensorNetworks2018,moritaGlobalOptimizationTensor2021,hommaNuclearNormRegularized2024}. Since optimization-based TRG methods require many QR deconposition and SVD operations at each coarse-graining step, reverse-mode differentiation becomes substantially more demanding. 
\begin{acknowledgments}
We thank Shinichiro Akiyama, Jutho Haegeman, Atsushi Ueda and Adwait Naravane for fruitful discussions.
Y.S. is supported by Graduate Program on Physics for the Universe (GP-PU), Tohoku University and JSPS KAKENHI Grant No. 25KJ0537.
Numerical calculations were performed on the supercomputer Pegasus under the Multidisciplinary Cooperative Research Program in CCS, University of Tsukuba.
\end{acknowledgments}
\appendix
\section{About normalization}
In practical TRG implementations, one must carefully avoid numerical overflow. A common strategy is to normalize the tensor at each coarse-graining step as
\begin{equation}\label{trT}
  \tilde{T}^{(i)} = \frac{T^{(i)}}{\mathrm{tTr}\left[T^{(i)}\right]} .
\end{equation}
where $\tilde{T^{(i)}}$ denotes the normalized tensor at $i$th step.
By such normalization, the free energy at $n$th iteration can be written as,
\begin{equation}
  \frac{\ln Z^{(n)}}{V^{(n)}}=\sum_{i=0}\frac{\ln{k^{(i)}}}{2^i}.
\end{equation}
where $k^{(i)}=\mathrm{tTr}\left[T^{(i)}\right]$ and $V^{(n)}=2^{n}$.
In this way, the first derivative of $\ln Z/V$ with respect to $\beta$ at the $n$th iteration is given by
\begin{equation}
  \frac{1}{V^{(n)}}\pdv{\ln Z^{(n)}}{\beta}
  =\sum_{i=0}^{n} \frac{1}{2^{i}} \frac{\dot{k}^{(i)}}{k^{(i)}},
\end{equation}
where
\begin{equation}
  \dot{k}^{(i)} = \pdv{k^{(i)}}{\beta}
  = \mathrm{tTr}\left[\dot{T}^{(i)}\right].
\end{equation}
In the next coarse-graining step, we have to use the derivative of normalized tensor $\tilde{T}^{(i)}$, and its derivative,
\begin{equation}
  \dot{\tilde{T}}^{(n)}
  = \frac{\dot{T}^{(n)}k^{(n)} - T^{(n)}\dot{k}^{(n)}}{\left(k^{(n)}\right)^2}.
\end{equation}
For the second derivative case, we obtain
\begin{equation}
  \ddot{\tilde{T}}^{(n)}=\frac{\ddot{T}^{(n)}\left(k^{(n)}\right)^2-2\dot{T}^{(n)}\dot{k}^{(n)}k^{(n)}-T^{(n)}\ddot{k}^{(n)}k^{(n)}+2T^{(n)}\left(\dot{k}^{(n)}\right)^2}{\left(k^{(n)}\right)^3},
\end{equation}
and
\begin{equation}
  \frac{1}{V^{(n)}}\frac{\partial^{2}\ln Z^{(n)}}{\partial\beta^{2}}=\sum_{i=0}^{n}\frac{1}{2^{i}}\left[\frac{\ddot{k}^{(i)}}{k^{(i)}}-\left(\frac{\dot{k}^{(i)}}{k^{(i)}}\right)^2\right].
  \label{eq:second_derivative_k}
\end{equation}

\section{Differentiation of SVD}\label{ap:SVD}
In this section, we write down the forward-mode AD rule for SVD used in HOTRG and BWTRG.
This section is based on Ref.~\cite{townsend2016differentiating}.
In these algorithms, the decomposition is applied to square matrices; hence, we restrict to
$A(\beta)\in\mathbb{R}^{n\times n}$.
Let $A(\beta)$ be a smooth function of a scalar parameter $\beta$, and assume we have a full SVD
$A=U\Sigma V^{T}$ with $U,V\in\mathbb{R}^{n\times n}$.
For simplicity, we assume that even numerically small singular values are treated as finite, so that the
corresponding singular vectors are nontrivial.
We denote
\begin{align}
  \dot{A} = \frac{dA}{d\beta},\qquad
  \ddot{A} = \frac{d^2A}{d\beta^2}.
\end{align}
We use the full SVD in the form
\begin{align}
  A = U \Sigma V^{T},
\end{align}
where $U,V\in\mathbb{R}^{n\times n}$ are orthogonal,
\begin{align}
  U^{T}U=I_n,\qquad V^{T}V=I_n,
\end{align}
and $\Sigma=\mathrm{diag}(\sigma_1,\ldots,\sigma_n)$ with $\sigma_i> 0$.

Let us introduce the matrix
\begin{align}
  P = U^{T}\dot{A}V.
\end{align}
Then the first derivative of the singular values is given by its diagonal part,
\begin{align}
  \dot{\sigma}_i = P_{ii},\qquad
  \dot{\Sigma}=\mathrm{diag}(\dot{\sigma}_1,\dots,\dot{\sigma}_n).
\end{align}
For the singular vectors, we parametrize the derivatives as
\begin{align}
  \dot{U}=U\Omega_U,\qquad \dot{V}=V\Omega_V,
\end{align}
where $\Omega_U,\Omega_V\in\mathbb{R}^{n\times n}$ satisfy the following conditions:
\begin{align}
  \Omega_U^{T}=-\Omega_U,\qquad \Omega_V^{T}=-\Omega_V.
\end{align}
For $i\neq j$, assuming $\sigma_i\neq\sigma_j$, we then have
\begin{align}
  (\Omega_U)_{ij}
  &=
  \frac{\sigma_j P_{ij}+\sigma_i P_{ji}}{\sigma_j^2-\sigma_i^2},
  \\
  (\Omega_V)_{ij}
  &=
  \frac{\sigma_i P_{ij}+\sigma_j P_{ji}}{\sigma_j^2-\sigma_i^2}.
\end{align}
In tensor network calculations, one often encounters degenerate singular values, for which the denominators
$\sigma_j^2-\sigma_i^2$ become zero and the SVD derivatives are ill defined.
Following Ref.~\cite{liaoDifferentiableProgrammingTensor2019}, we adopt a Lorentzian regularization by defining
\begin{align}
  \frac{1}{\sigma_j^2-\sigma_i^2}\ \longrightarrow\ \frac{\sigma_j^2-\sigma_i^2}{(\sigma_j^2-\sigma_i^2)^2+\eta},
  \qquad \eta>0.
\end{align}
The forward-mode derivatives are then obtained as
\begin{align}
  \dot{U}=U\Omega_U,\qquad
  \dot{V}=V\Omega_V,\qquad
  \dot{\Sigma}=\mathrm{diag}(P).
\end{align}

Even when a truncated SVD is used in TRG, the AD must be formulated at the level of the full SVD. This is because the derivatives of the retained vectors mathematically depend on the singular values and vectors of the discarded sector. While Ref.~\cite{liaoDifferentiableProgrammingTensor2019} neglected this contribution, Ref.~\cite{francuzStableEfficientDifferentiation2025} pointed out that the truncated subspace must be correctly included in the derivation.
The second-order derivatives are obtained by differentiating the first-order relations once more with respect to $\beta$.
\section{Computational cost}
\subsection{General case}\label{ap:general_case}
In this section, we provide details of a contraction-tree-based differentiation method. 
A contraction-tree~\cite{evenblyImprovingEfficiencyVariational2014} is a binary tree that specifies the contraction order of a tensor networks.
While contraction-tree-based reverse mode automatic differentiation algorithms are described in Ref.~\cite{gorodetskyReversemodeDifferentiationArbitrary2022}, our method provides a general framework to compute arbitrary $k$th derivatives of tensor networks by forward-mode automatic differentiation.

\begin{figure}[tbp]
  \centering
  \begin{subfigure}[b]{0.48\textwidth}
    \centering
    \includegraphics[width=\textwidth]{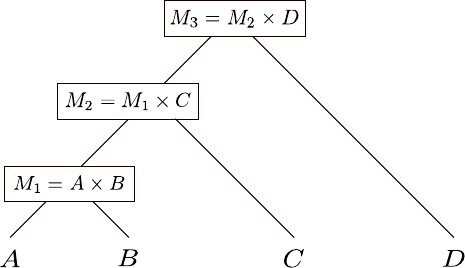}
    \caption{ }
    \label{fig:contraction_tree}
  \end{subfigure}
  \hfill
  \begin{subfigure}[b]{0.48\textwidth}
    \centering
    \includegraphics[width=\textwidth]{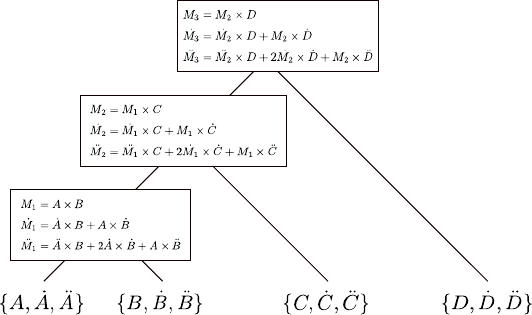}
    \caption{ }
    \label{fig:contraction_tree2}
  \end{subfigure}
  \caption{(a) The contraction-tree for Eq.~\eqref{eq:cont_tree}. (b) Modified contraction-tree for evaluating derivatives up to second order.}
  \label{fig:combined_trees}
\end{figure}

Let us begin with an example. Consider the following contraction, where the dotted circles indicate contractions between pairs of tensors
\begin{equation}\label{eq:cont_tree}
  \vcenter{\hbox{\includegraphics[scale=1.0]{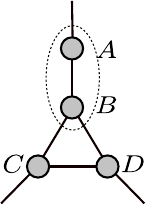}}}\,\rightarrow\,\vcenter{\hbox{\includegraphics[scale=0.9]{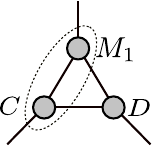}}}\,\rightarrow\,\vcenter{\hbox{\includegraphics[scale=0.9]{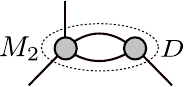}}}\,\rightarrow\,\vcenter{\hbox{\includegraphics[scale=0.9]{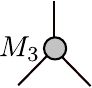}}}
\end{equation}
In this contraction, we evaluate the intermediate tensors and their derivatives in a single forward pass along the contraction-tree. Let us denote the contraction of two tensors $A$ and $B$ by $\mathrm{contract}(A,B)$. Then, the contraction-tree-based differentiation can be written as
\begin{align}
  \textbf{Input:}\quad
  &(A,\dot{A},\ddot{A}),\ (B,\dot{B},\ddot{B}),\ (C,\dot{C},\ddot{C}),\ (D,\dot{D},\ddot{D})\nonumber\\
  M_1 &= \text{contract}(A,B)\label{eq:M1}\\
  \dot{M}_1 &= \text{contract}(\dot{A},B)+\text{contract}(A,\dot{B})\label{eq:dM1}\\
  \ddot{M}_1 &= \text{contract}(\ddot{A},B)+2\,\text{contract}(\dot{A},\dot{B})+\text{contract}(A,\ddot{B})\label{eq:d2M1}\\
  M_2 &= \text{contract}(M_1,C)\\
  \dot{M}_2 &= \text{contract}(\dot{M}_1,C)+\text{contract}(M_1,\dot{C})\\
  \ddot{M}_2 &= \text{contract}(\ddot{M}_1,C)+2\,\text{contract}(\dot{M}_1,\dot{C})+\text{contract}(M_1,\ddot{C})\\
  M_3 &= \text{contract}(M_2,D)\\
  \dot{M}_3 &= \text{contract}(\dot{M}_2,D)+\text{contract}(M_2,\dot{D})\\
  \ddot{M}_3 &= \text{contract}(\ddot{M}_2,D)+2\,\text{contract}(\dot{M}_2,\dot{D})+\text{contract}(M_2,\ddot{D})\\
  \textbf{Output:}\quad
  &(M_3,\dot{M}_3,\ddot{M}_3)\nonumber
\end{align}
In the above equations, the intermediate tensors are differentiated at each contraction step.
A schematic picture of the contraction-tree for Eq.~\eqref{eq:cont_tree} is shown in Fig.~\ref{fig:contraction_tree}.
To evaluate derivatives of this network up to second order, the contraction-tree is modified as illustrated in Fig.~\ref{fig:contraction_tree2}.
The total cost of evaluating derivatives up to second order is exactly 6 times the usual cost.

It is straightforward to derive general formula. If the tensor network $G$ is initially represented by a set of $n$ tensors,
\begin{equation}
  \mathcal{G}^{(0)} = \left\{ A_1^{(0)}, A_2^{(0)}, \dots, A_n^{(0)} \right\},
\end{equation}
and assume that each tensor has a parameter dependence on $\beta$.
The full contraction of this tensor network is denoted as
\begin{equation}
  G= \text{contract}(A_1^{(0)}, A_2^{(0)}, \dots, A_n^{(0)}).
\end{equation}
can be performed recursively in $n-1$ steps.
At the $i$th step ($i=1, \dots, n-1$), the system transitions from the network $\mathcal{G}^{(i-1)}$ to a new set of $n-i$ tensors:
\begin{equation}
  \mathcal{G}^{(i)} = \left\{ A_1^{(i)}, A_2^{(i)}, \dots, A_{n-i}^{(i)} \right\}.
\end{equation}
This transition is defined by selecting two tensors, $A_p^{(i-1)}$ and $A_q^{(i-1)}$, from the previous network $\mathcal{G}^{(i-1)}$ and contracting them to form the new tensor $A_{n-i}^{(i)}$:
\begin{equation}
  A_{n-i}^{(i)} = \text{contract}\left( A_p^{(i-1)}, A_q^{(i-1)} \right).
\end{equation}
The remaining $n-i-1$ tensors in $\mathcal{G}^{(i)}$ correspond to the unselected tensors carried over from $\mathcal{G}^{(i-1)}$.
Now let us introduce the notation $A[k]$ to represent the $k$th derivatives of a tensor $A$, formally defined as
\begin{equation}
  A[k] \equiv \overset{\overbrace{\cdot\cdot\cdot\cdot\cdot\cdot\cdot}^{k}}{A}.
\end{equation}
Assume we have access to the derivatives of the initial tensor set $\mathcal{G}^{(0)}$ up to the $k$th order.
Using the above notation, we define the set of $m$th derivatives, denoted as $\mathcal{G}^{(0)}[m]$ as follows
\begin{equation}
  \begin{split}
    \mathcal{G}^{(0)}[1] &= \left\{ A_1^{(0)}[1], A_2^{(0)}[1], \dots, A_n^{(0)}[1] \right\}\\
    \mathcal{G}^{(0)}[2] &= \left\{ A_1^{(0)}[2], A_2^{(0)}[2], \dots, A_n^{(0)}[2] \right\}\\
    &\quad \vdots \\
    \mathcal{G}^{(0)}[k] &= \left\{ A_1^{(0)}[k], A_2^{(0)}[k], \dots, A_n^{(0)}[k] \right\},
  \end{split}
\end{equation}
where $A_j^{(0)}[m]$ corresponds to the $m$th derivative of the $j$th tensor in the initial set.
At each contraction step, calculating the derivatives of the new tensor $A_{n-i}^{(i)}$ follows the general Leibniz rule:
\begin{align}
  A_{n-i}^{(i)}[1] &= \text{contract}\left( A_p^{(i-1)}[1], A_q^{(i-1)}[0] \right) + \text{contract}\left( A_p^{(i-1)}[0], A_q^{(i-1)}[1] \right), \\
  A_{n-i}^{(i)}[2] &= \text{contract}\left( A_p^{(i-1)}[2], A_q^{(i-1)}[0] \right) + 2\,\text{contract}\left( A_p^{(i-1)}[1], A_q^{(i-1)}[1] \right) + \text{contract}\left( A_p^{(i-1)}[0], A_q^{(i-1)}[2] \right), \\
  &\quad \vdots \notag \\
  A_{n-i}^{(i)}[k] &= \sum_{m=0}^{k} \binom{k}{m} \text{contract}\left( A_p^{(i-1)}[m], A_q^{(i-1)}[k-m] \right).
\end{align}
Consequently, after completing all $n-1$ contraction steps, the network reduces to a single final tensor $A_1^{(n-1)}$.
Thus, we obtain the derivatives of the network $G$ for all orders up to $k$,
\begin{equation}
  G[m] = A_1^{(n-1)}[m], \qquad m = 0, 1, \dots, k.
\end{equation}
The total cost of evaluating up to $k$th derivatives is exactly $\frac{(k+1)(k+2)}{2}$ times the original contraction.

\subsection{Linear scaling algorithm}
The contraction-tree-based algorithm is itself very efficient in terms of the computational cost, however, a simple modification can be considered, which results in the total computational cost to be $2k+1$ times original contraction. Consider the second derivative case. We consider contraction in Eqs.~\eqref{eq:M1}--\eqref{eq:d2M1}.
Let $C$ be the total cost of evaluating Eq.~\eqref{eq:M1}.
The total cost of evaluating contraction in Eqs.~\eqref{eq:M1}--\eqref{eq:d2M1} is $6C$.
The key idea is to take combination of them, as
\begin{align}
  &M_1=\mathrm{contract}(A,B)\\
  &S_+=\mathrm{contract}(A+\dot{A},B+\dot{B}),\qquad S_-=\mathrm{contract}(A-\dot{A},B-\dot{B})\\
  &T_+=\mathrm{contract}(A+\ddot{A},B+\ddot{B}),\qquad T_-=\mathrm{contract}(A-\ddot{A},B-\ddot{B}).
\end{align}
We can construct $\dot{M}_1$ and $\ddot{M}_1$ from these tensors:
\begin{align}
  \dot{M}_1&=\frac{S_+-S_-}{2}\\
  \ddot{M}_1&=\frac{T_+-T_-}{2}+2\left(\frac{S_++S_-}{2}-M_1\right).
\end{align}
In this scheme, the total cost of the matrix multiplication is $5C$, slightly smaller than $6C$.

This can be generalized in terms of the Taylor-mode AD~\cite{bettencourt2019taylormode}. Suppose we want to evaluate up to $k$th derivatives of $M_1$.
We define Taylor mode for $A$ and $B$,
\begin{align}
  A(t) &= \sum_{m=0}^{k} \frac{A[m]}{m!}\, t^{m},\\
  B(t) &= \sum_{m=0}^{k} \frac{B[m]}{m!}\, t^{m}.
\end{align}
Then the product
\begin{align}
  W(t)= \mathrm{contract}(A(t),B(t))
\end{align}
is written as
\begin{align}
  W(t)=\sum_{m=0}^{k}\frac{M_1[m]}{m!}t^{m}+\sum_{m=k+1}^{2k}\frac{R_1[m]}{m!}t^{m}.
\end{align}
Thus, $W(t)$ serves as a generating function for $\{M_1[m]\}_{m=0}^{k}$, with additional coefficients $\{R_1[m]\}_{m=k+1}^{2k}$ that are not used in the computation of derivatives up to the order of $k$.
Since $W(t)$ is a polynomial of degree at most $2k$, it can be uniquely reconstructed by polynomial interpolation by choosing $2k+1$ distinct points.
In this scheme, the total cost of the matrix-multiplications is $(2k+1)C$, since we need to compute distinct points of $W$ at $2k+1$ points. Note that this interpolation scheme introduces an additional overhead from tensor additions and scalar multiplications, although the number of dominant contractions is reduced, which could be even slower than contraction-tree-based differentiation at small $k$.

\bibliography{TRG,for_this}
\end{document}